\documentstyle[aps,prd]{revtex}
\begin{document}

\title{Exact inhomogeneous cosmologies whose source is a radiation-matter mixture
with consistent thermodynamics.}

\author{Roberto A. Sussman$^\dagger$ and Diego
Pav\'on$^\ddagger$}
\address
{$^\dagger$
Instituto  de Ciencias Nucleares
UNAM, Apartado Postal 70-543,
\\ M\'exico DF, 04510, M\'exico.\\
$^\ddagger$ Departamento de F\'{\i}sica, Facultad de Ciencias,
Universidad Aut\'onoma de Barcelona, \\
E- 08193 Bellaterra, Espa\~na.}

\maketitle
\begin{abstract}
We derive a new class of exact solutions of Einstein's equations providing a
physically plausible hydrodynamical description of cosmological matter in the
radiative era, between nucleosynthesis and decoupling. The solutions are
characterized by the Lema\^{\i}tre-Tolman-Bondi metric with a viscous fluid
source, subjected to the following conditions: (a) the equilibrium state
variables satisfy the equation of state of a mixture of an ultra-relativistic
and a non-relativistic ideal gases, where the internal energy of the latter
has been neglected, (b) the particle numbers of the mixture components are
independently conserved, (c) the viscous stress is consistent with the
transport equation and entropy balance law of Extended Irreversible
Thermodynamics, with the coefficient of shear viscosity provided by Kinetic
Theory. The fulfilment of (a), (b) and (c) restricts initial conditions in
terms of an initial value function, $\Delta_i^{(s)}$, which in the limit of
small density contrasts becomes the average of spatial gradients of the
fluctuations of photon entropy per baryon in the initial hypersurface. For
$\Delta_i^{(s)}\ne 0$ and choosing the phenomenological coefficients of the
``radiative gas'' model, we have an interactive photon-baryon mixture under
local thermal equilibrium, with radiation dominance and temperatures
characteristic of the radiative era ($10^6\hbox{K} > T > 10^3\hbox{K}$).
Constraints on the observed anisotropy of the microwave cosmic radiation and
the condition that decoupling occurs at $T=T_{_D}\approx 4\times 10^3$ K
yield an estimated value: $|\Delta_i^{(s)}|\approx 10^{-8}$ which can be associated
with a bound on promordial entropy fluctuations. The Jeans mass
at decoupling is of the same order of magnitude as that of baryon dominated
perturbation models ($\approx 10^{16} M_\odot$).
\end{abstract}
\section {Introduction}

The radiative era of cosmic evolution comprises the period from the end of
primeval nucleosynthesis to the decoupling of matter and radiation
(see refs \cite{sw} to \cite{lan}). A gross description of cosmological
matter sources in
this period is given by an interactive mixture of ideal relativistic and
non relativistic
gases (``radiation'' and ``matter'') in local thermal equilibrium (LTE).

The standard approach to this type of matter source is either a FLRW
spacetime whith equilibrium Kinetic Theory
distributions\cite{kotu}, \cite{pad}, \cite{bern},
gauge invariant perturbations on a FLRW
background \cite{bor}, \cite{kotu}, \cite{pad}, \cite{peeb}, 
\cite{kosa}, \cite{efsta}, and various
types of hydrodynamical models, \cite{ct}, \cite{rdm}, \cite{pm} which, in
general, fail to
incorporate a physically plausible description of the interaction between
matter and
radiation. Even if we argue that the universe is ``almost FLRW'' or
``almost in thermal
equilibrium'', the small deviations from equilibrium are extremely important,
\cite{sw}, \cite{wei}, \cite{bor}, \cite{kotu}, \cite{pad},
\cite{peeb}, \cite{kosa}, to account for
most interesting phenomena of cosmic evolution: nucleosynthesis, structure 
formation, abundance of relic gases, etc. Models with perfect fluid 
sources, whether
hydrodynamic, \cite{ct}, \cite{rdm}, or based on Kinetic Theory, \cite{emt},
necessarily assume a
quasi-static adiabatic and reversible evolution and thus, fail to
incorporate into the
resulting picture even small deviations from equilibrium.

Dissipative sources have been incorporated numerically within a purely FLRW
geometry \cite{hs} or following a perturbative approach \cite{matr}. However,
the literature
still lacks an alternative hydrodynamical treatment, based on inhomogeneous
exact
solutions of Einstein's equations with dissipative sources and fully
complying with the
thermodynamics of a radiative gas within a transient regime. Ideally, such
exact models
should include all dissipative agents (heat flux, bulk and shear viscosity)
and should be
consistent with the theoretical framework of Extended Irreversible
Thermodynamics (see
refs \cite{matr} to \cite{schweitzer}), thus satisfying suitable transport
equations
complying with causality, with phenomenological coefficients given by
Kinetic Theory for
this type of source. Since this general treatment would be mathematically
untractable, we
aim at the best possible approach based on exact solutions of Einstein's
equations.
Therefore, we have made the following simplifying assumptions: (a) the
matter source is a
fluid with shear viscosity but without heat conduction nor bulk viscosity,
(b) the
equilibrium state variables satisfy the equation of state of a mixture of
relativistic
and non-relativistic ideal gases, where the internal energy and pressure of
the latter
have been neglected, (c) the particle numbers of each mixture components is
independently
conserved, (d) we exclude dark matter and/or exotic particles and assume
instead a tight
coupling between photons (radiation) and baryons and electrons (matter),
hence there is a common temperature for the mixture
(LTE), while the
microscopical interaction models are the various processes of radiative
transfer, \cite{wei}, \cite{pad}, \cite{peeb}, \cite{jcl}, 
\cite{uwi}, \cite{dd1}, \cite{dd2}:
Thomson
scattering, brehmstrallung, free-free absortion, etc. Although this type of
interactions
involve mostly photons and electrons, the dynamics of the matter component
is governed by
the baryons since the latter provide most of the rest mass content of
non-relativistic
matter (without dark matter).

Restrictions (b) and (c) are easy to
justify: since the ratio of photons to baryons is such a large number
($\approx 10^9$), we can truly ignore the pressure of non-relativistic
matter. Also, after nucleosynthesis, in the temperature range $T<10^6$,
matter creation and anhilitation processes balance each other and 
effectively cease to be dynamically important \cite{kotu}, \cite{pad}. 
On the other hand, the lack of
heat conduction (restriction (a)) is more difficult to justify.
It can be associated with an adiabatic (zero heat flux) but still irreversible
evolution (nonzero viscosity), and can be a reasonable approximation on
specific conditions. For example, for a radiative gas at higher temperatures
shear viscosity dominates over heat conduction, but the latter becomes
significant as the mixture cools\cite{wei}. The lack of bulk viscosity is a
better
approximation: it is negligible for a radiative gas in the temperature range
$10^3\,\hbox{K} < T < 10^6\,\hbox{K}$ that we are
interested, \cite{wei}, \cite{matr}, \cite{wjs}, \cite{jcl}, \cite{uwi}, 
becoming important for
higher temperatures (the mid relativistic regime where
$k_{_{B}}T\approx mc^2$ \cite{matr}, \cite{zim}, \cite{schweitzer}).  
However, we accept that
ignoring these dissipative fluxes weakens the scope and validity of the
models, but we
argue that this is compensated by the simplification of the field
equations, leading to
exact forms for the equilibrium state variables and shear viscosity that still
satisfy (under the restrictions mentioned) thermodynamically consistent
relations.

The models we present are based on the spherically
symmetric Lema\^{\i}tre-Tolman-Bondi metrics, usually associated with dust
sources \cite{ksmh}, \cite{krs}. However, this metric is compatible with a
comoving fluid
source with zero heat flux but with anisotropic stresses, which we describe
as shear
viscosity. Obviously, the lack of heat flux and 4-acceleration
necessarily implies a very special shear viscous tensor whose
divergence exactly balances the nonzero spatial gradient of the equilibrium
pressure. Considering this metric and this source, we impose on the
equilibrium state variables the equation of state for a mixture of ideal
gases (under the restriction (b)). The field equations can be
solved up to a quadrature, without having to make any assumption on
the form of the shear viscous pressure. The latter, as well as all
equilibrium state variables can be determined from the solution of
the quadrature, up to two initial value functions that can be
identified with the initial energy densities of the matter and
radiation components. We consider only the case that would be equivalent to
spacelike sections of zero curvature. A generalization of this class of
exact solutions to the more general Szekeres-Szafron metrics admitting no
isometries
has been published recently \cite{susstr}, while the study of a
non-relativistic ideal gas is considered in \cite{suss98}.

Once the field equations have been integrated, we define a set of
initial value functions that gauge the deviation from homogeneity of
the average of initial density contrasts. The terms
involving various gradients of metric functions can be given in terms of
these gauges, so that in the limit when the latter vanish a FLRW spacetime can
always be obtained as the homogeneous (and reversible)
subcase. In section VII we derive the conditions that the models must
satisfy in order to be consistent with the theoretical framework
of Irreversible Extended Thermodynamics, in the case where shear viscosity is
the only dissipative agent and the coefficient of shear viscosity is that
given by Kinetic Theory for the radiative
gas \cite{wei}, \cite{jcl}, \cite{uwi}, \cite{dd1}, \cite{dd2}, 
\cite{schweitzer}. This leads to an
entropy balance law and a suitable transport equation for shear viscosity
that is
satisfied for a specific functional form of the relaxation time. Conditions
are given so
that the latter quantity behaves as a relaxation parameter for an interactive
cosmological mixture of matter and radiation. These conditions of
thermodynamical
consistency are then explicitly tested on the models, leading to a set of
restrictions on
the initial conditions (the latter given in terms of the gauges of
initial density constrasts). The most relevant result is
that thermodynamical consistency constrains an initial
value adimensional function, $\Delta_i^{(s)}$, which in the limit of
small density contrasts is approximately the average
gradient of the photon entropy per baryon along the initial hypersurface
$t=t_i$. An analogy is provided with the theory of perturbations on a FLRW
background, whereby $\Delta_i^{(s)}=0$ is formally analogous
to the definition of initialy adiabatic perturbations in the
sincronous gauge \cite{bor}, \cite{pad}, \cite{kosa}, \cite{efsta}. The
constraints on the
observed anisotropy of the microwave cosmic background, as well as
the condition that decopling occurs at $T=T_{_{D}}\approx
4\times 10^3\hbox{K}$, leads to the estimated value
$|\Delta_i^{(s)}|\approx 10^{-8}$. Since initial conditions of the radiative
era should be traced to previous periods of cosmic evolution, this constraint
can be related to maximal bounds on entropy fluctuations in primordial
perturbations. Finally, we compute the Jeans mass associated
with the thermodynamically consistent models, leading to a
value similar to that obtained for baryon
dominated perturbation models: $M_{_{J}}\approx 10^{16}
M_\odot$.

\section{Interacting mixture of radiation and non-relativistic matter}

A radiation-matter mixture can be described by a mixture of two ideal
gases: one
an ultra-relativistic gas of massless particles, the other a
non-relativistic ideal monatomic
gas with $m$ being the mass of the particles. This is characterized 
by the total matter energy, $\rho$ and pressure $p$

$$\rho= mc^2n^{(m)}+{3\over
2}n^{(m)}k_{_B}T^{(m)}+3n^{(r)}k_{_B}T^{(r)}, \eqno(1a)$$
$$p=
n^{(m)}k_{_B}T^{(m)}+n^{(r)}k_{_B}T^{(r)}, \eqno(1b)$$

\noindent
where $k_{_B}$ is Boltzmann's constant and $n,T$ are particle number
densities and
temperatures of the two components, distiguished by the superindices $(m)$
(``matter'') and
$(r)$ (``radiation''). If there is local thermal equilibrium (LTE) between
the components, the
latter interact and evolve with the same temperature: $T^{(r)}=T^{(m)}=T$.
If the components
are decoupled, each gas evolves with a different temperature.

Assuming LTE, if $n^{(m)}\ll n^{(r)}$, but the ratio $mc^2/k_{_B}T$ is not
negligible, then
equations (1) can be approximated by

$$\rho\approx mc^2n^{(m)}+3n^{(r)}k_{_B}T, \eqno(2a) $$
$$p\approx n^{(r)}k_{_B}T , \eqno(2b) $$

\noindent
an equation of state describing a radiation dominated mixture in which the
presence of
non-relativistic matter is dynamically important. If we assume
non-relativistic matter to be
made up of baryons (with
$m$ being a protonic mass) and since the ratio of baryons to photons
$n^{(m)}/n^{(r)}\approx
10^{-9}$ is a small number, the equation of state (2) is a reasonable
approximation in the
temperature range $10^3 \leq T
\leq 10^6 \,
\mbox{K}$, characteristic of the ``radiative era" from the end of
nucleosynthesis to the
transition between radiation to matter dominance, including the
recombination and decoupling
eras. At such temperatures, it is also safe to assume,
\cite{sw},\cite{na},\cite{bor},\cite{kotu},\cite{pad} that electrons and
photons interact mostly
through Thomson scattering but creation and annihilation processes
(bremsstrahlung and
free-free absorption) roughly compensated one another so that particle
number densities of
the components of the mixture satisfy independent conservation laws. Once
the decoupling of
the matter-radiation mixture takes place at about
$T\approx 4\times 10^3$ K, the assumption of LTE is no longer valid and
interaction between
components ceases. Equation of state (1) can also be
approximated by a form similar (2) with the internal energy of radiation taking
approximately the Stefan-Boltzmann law:
$\rho^{(r)}=a_{B}T^4$, where $a_{B}$ denotes the radiation constant.
However,  out of thermal equilibrium the Steffan-Boltzmann law is incompatible
with the ideal gas equation of state.

Having in mind the conditions justifying (2), we will
describe a matter-radiation mixture evolving along adiabatic but irreversible
processes by the fluid tensor

$$T^{ab}= \rho u^au^b+ph^{ab}+\Pi^{ab} \, , \eqno(3)$$ 
$$h^{ab}= c^{-2}u^a u^b+g^{ab},\qquad
u_a\Pi^{ab}= 0, \qquad \Pi^a\,_a = 0,$$

\noindent
where: $\rho,p$ satisfy (2), $u^a$ is the 4-velocity shared by radiation
and matter,
$\Pi^{ab}$ is the shear viscous pressure tensor (a symmetric traceless
tensor) which arises because of the matter-radiation interaction,
and particle number densities satisfy the conservation laws

$$(n^{(m)}u^a)_{;a}= 0,\qquad (n^{(r)}u^a)_{;a}= 0. \eqno(4)$$

\noindent
As mentioned previously, bulk viscosity is negligible within the 
temperature range we are interested in
\cite{jcl},\cite{wei},\cite{uwi},\cite{dd1},\cite{dd2},\cite{zim}, while
even if neglection of heat conduction can be justified for relativistic
temperatures\cite{wei}, it does weaken the scope of the models. However, this
restriction is compensated by the obtention of exact solutions that are still
thermodynamically consistent.

\section{The Lema\^{\i}tre-Tolman-Bondi metrics}

Consider (3) as the source of the ``Lema\^{\i}tre-Tolman-Bondi'' (LTB)
metric ansatz,
usually associated with spherically symmetric Lema\^{\i}tre-Tolman-Bondi
dust solutions
\cite{ksmh},\cite{krs}

$$ds^2= -c^2dt^2+{Y'^2\over{1-F}}\,dr^2 +Y^2\left(d\theta^2+\sin^2
(\theta) d\phi^2
\right)\, , \eqno(5)
$$

\noindent
where $Y= Y(t,r)$, $F= F(r)$,  and a prime denotes partial
derivative with  respect
$r$. Just as in the LTB dust solutions, we assume the coordinates in (5) to
be comoving and
the 4-velocity of the fluid source to be $u^a= c\,\delta^a\,_t $, a
geodesic vector field,
since $\dot u_a\equiv u_{a;b}u^b= 0$. Other kinematic invariants associated
with (5) are the
scalar expansion $\Theta\equiv u^a\,_{;a}$, and the shear tensor
$\sigma_{ab}\equiv
u_{(a;b)}-(\Theta/3)h_{ab}$, given in the coordinates of (5) by

$$\Theta= {\dot Y'\over{Y'}}+{{2\dot Y}\over Y} \, , \eqno(6a)$$

$$\sigma^a\,_b= {\bf{\hbox{diag}}}\left[0,-2\sigma,\sigma,\sigma \right],
\qquad\qquad
\sigma\equiv{1\over 3}\left({\dot Y\over Y}-{\dot Y'\over Y'} \right)
\, , \eqno(6b) $$

\noindent
while the most general form of $\Pi^a\,_b$ for the metric
(5) is given by

$$\Pi^a\,_b= {\bf{\hbox{diag}}}\left[0,-2P,P,P \right], \eqno(6c) $$

\noindent
where $\dot Y\equiv u^aY_{,a}= Y_{,t}$ and $P= P(t,r)$ is an arbitrary
function. Notice that a comoving and
non-accelerating 4-velocity does not imply $p'=0$, as in the perfect fluid
case ($G^\theta\,_\theta-G^r\,_r=0$). As revealed by the momentum balance law:
$h_{ca}T^{ab}\,_{;b}=0$, applied to the viscous fluid source (3), we have 

$$h_a\,^b(p_{,b}+\Pi_{bc;d}h^{cd})=0\qquad
\Rightarrow\qquad \left( p-2P\right){}'+6P\frac{Y'}{Y}=0\eqno(7a)$$

\noindent
showing how the divergence of the shear viscous tensor exactly balances the
nonzero pressure gradient. The energy balance, $u_aT^{ab}\,_{;b}=0$, is given
by  

$$\dot\rho+(\rho+p)\Theta +\sigma_{ab}\Pi^{ab}=0\qquad\Rightarrow\qquad  \dot
p +\frac{4}{3}\Theta p+6\sigma P=0\eqno(7b)$$

\noindent
illustrating how the term $\sigma_{ab}\Pi^{ab}=6\sigma P$ can be understood as
an interaction term responsible for local energy exchange between matter and
radiation. 

Integration of the conservation laws (4) for (5)
yields

$$n^{(m)}= n_i^{(m)}\left(\frac{Y_i}{Y}\right)^3\,\frac{Y'_i/Y_i}{Y'/Y} ,
\qquad n^{(r)}=
n_i^{(r)}\left(\frac{Y_i}{Y}\right)^3\,\frac{Y'_i/Y_i}{Y'/Y}\, , \eqno(8) $$

\noindent
where $n_i^{(m)},\,n_i^{(r)}$ depend only on $r$ and are the particle
number  densities of
non-relativistic matter and radiation, evaluated along a suitable initial
hypersurface labeled
by $t= t_i$. The subindex ``$_i$'' affixed to any quantity, as $Y_i$, will
denote henceforth
initial value functions (functions of $t,r$ evaluated along $t= t_i$). It is
important to state
that our initial conditions do not refere to present cosmic time 
(usually labeled
as $t=t_0$), and so we will not use the subindex ``$_0$''.

The spherically symmetric LTB metrics (5) are contained within a larger
class of more general metrics (the Szekeres-Szafron metrics \cite{ksmh},
\cite{krs}), admitting in general no isometries. The integration of the field
equations for (5), given a source (3) satisfying (2), is examined in the next
section. For the case of more general Szekeres-Szafron metrics,
see\cite{susstr}.

\section{Integration of the field equations.}

Einstein's field equations for (5) and (3) are

$$\kappa \rho=-\frac{\left[Y\left( \dot Y^2+Fc^2\right)
\right]{}'}{Y^2Y'}=-G^t\,_t \, , \eqno(9a)$$
$$\kappa p=-\frac{\left[Y\left( \dot Y^2+Fc^2\right) +2Y^2\ddot
Y\right]{}'}{3Y^2Y'}=\frac{1}{3}\left(2G^\theta\,_\theta+G^r\,_r
\right) \, ,
\eqno(9b)$$
$$\kappa P=\frac{Y}{6Y'}\left[\frac{Y\left( \dot Y^2+Fc^2\right) +2Y^2\ddot
Y}{Y^3} \right]{}'= \frac{1}{3}\left(G^\theta\,_\theta-G^r\,_r
\right) \, , \eqno(9c)$$

\noindent where $\kappa\equiv 8\pi G/c^2$. Imposing on (9a) and (9b) the
equation of state (2), using (8) and integrating with respect
to $r$ yields the following constraint

$$2Y(\dot Y^2+Fc^2)+Y^2\ddot Y-\kappa\, mc^2\int{n_i^{^{(m)}}Y_i^2Y'_idr}=
\lambda(t) \, ,\eqno(10) $$

\noindent where $\lambda(t)$ is an arbitrary
integration function. It is important to remark that (10) follows only
from (9a) and (9b) without involving (9c), {\it ie} it was
not necessary to make any assumption regarding the form of $P$ in order to
obtain (10). A second integration of the field equations necessarily requires
setting $\lambda(t)= 0$ in (10), leading to

$$\dot Y^2= {{\kappa}\over{Y}}\left[M+W\left ({Y_i\over Y} \right)
\right]-Fc^2 , \eqno(11) $$

\noindent
where

$$ M= \int{\rho_i^{(m)}Y_i^2Y'_idr},\qquad \rho_i^{(m)}\equiv mc^2
n_i^{^{(m)}} \, , \eqno(12a)$$
$$W= \int{\rho_i^{(r)}Y_i^2Y'_idr},\qquad \rho_i^{(r)}\equiv
3n_i^{^{(r)}}k_{_B}T_i \, ,
\eqno(12b)$$

\noindent
so that $\rho_i^{(m)},\,\rho_i^{(r)} $ respectively define the initial
densities of
the non-relativistic and  relativistic components of the mixture.

In the remaining of the paper we restrict ourselves to $F= 0$, similar to
the choice of spacelike sections of zero curvature in FLRW geometry, leaving
the case $F\ne 0$ for a future analysis. An explicit integral of (11) in this
case is given by

$$\frac{3}{2}\sqrt{\mu} \,(t-t_i)=
\sqrt{y+\epsilon}\left(y-2\epsilon
\right)-
\sqrt{1+\epsilon}\left(1-2\epsilon\right) \, , \eqno(13a)$$

\noindent where
$$\mu\equiv  \frac{\kappa M}{Y_i^3},\qquad \epsilon\equiv
 \frac{W}{M},\qquad
y\equiv\frac{Y}{Y_i}\, , \eqno(13b)$$

\noindent
It is possible to invert (13a), thus obtaining $y=y(t,r)$ as a
complicated, but closed analytic form, where the $r$ dependence is contained
in the functions $\mu,\,\epsilon$ appearing in (13b). However
it turns out to be more convenient to use (11) and (13) to simplify the field
equations and radial gradients of $y$ in order to express all state and
geometric variables in terms of $y$ and suitable initial value functions
related to those of (12) and (13b).

\section{The state variables.}

From (8) and (11)-(13) it is possible to obtain the state variables
$n^{(m)},\,n^{(r)},\,T,\rho^{(m)},\,\rho^{(r)},\,p,\,P$. However, before
doing so it is useful to define the averaged initial densities

$$\left\langle \rho_i^{(m)}\right\rangle\equiv \frac{\int{\rho_i^{(m)}
d(Y_i^3)}}{Y_i^3}= \frac{3M}{Y_i^3},
\qquad \left\langle \rho_i^{(r)}\right\rangle \equiv
\frac{\int{\rho_i^{(r)} d(Y_0^3)}}{Y_i^3}= \frac{3W}{Y_i^3}\, , \eqno(14) $$

\noindent averaged over the volume
$Y_i^3$. Since the
solutions  allow for an arbitrary re-scaling of the radial coordinate,
without loss of generality we can select $Y_i= rR_i$, where $R_i$ is a
characteristic constant length scale.
Therefore, the volume $Y_i^3$, evaluated from the symmetry center, $r= 0$,
to an arbitrary
fluid layer $r$, can be characterized invariantly as the volume of the
orbits of the rotation group
$\hbox{SO}(3)$ in the hypersurface $t= t_i$. In the Newtonian limit, the
distance $Y_i$
becomes the radius of the circular keplerian orbit in the field of (5).

Together with the averaged initial densities, we shall define the quantities
$\Delta_i^{(m)},\,
\Delta_i^{(r)}$ given by

$$\Delta_i^{^{(m)}}= \frac{\int{[\rho_i^{(m)}]'Y_0^3 dr}}{3M}=
\frac{\rho_i^{(m)}}{\left\langle
\rho_i^{(m)}\right\rangle}  -1\quad \Rightarrow\quad \rho_i^{(m)}=\left\langle
\rho_i^{(m)}\right\rangle\left[1+\Delta_i^{^{(m)}}\right], \eqno(15a)$$
$$\Delta_i^{^{(r)}}= \frac{\int{[\rho_i^{(r)}]'Y_i^3 dr}}{3W}=
\frac{\rho_i^{(r)}}{\left\langle
\rho_i^{(r)}\right\rangle}  -1\quad \Rightarrow\quad \rho_i^{(r)}=\left\langle
\rho_i^{(r)}\right\rangle\left[1+\Delta_i^{^{(r)}}\right], \eqno(15b)$$

\noindent whose interpretation as effective initial density contrasts is
discussed
in the following section. Using (14) and (15) we can re-write
$\mu,\epsilon$ in (13b) as

$$\mu= \frac{\kappa}{3}\left\langle
\rho_i^{(m)}\right\rangle=\frac{\kappa\rho_i^{(m)}
}{3(1+\Delta_i^{^{(m)}})},\qquad
\epsilon=\frac{\left\langle \rho_i^{(r)}
\right\rangle}{\left\langle
\rho_i^{(m)}\right\rangle}=\frac{\rho_i^{(r)}}{\rho_i^{(m)}}\,
\frac{1+\Delta_i^{^{(m)}}}{1+\Delta_i^{^{(r)}}}.\eqno(16)$$

\noindent
The state variables $n^{(m)},\,n^{(r)},\,\rho,\,p,\,P$ now follow by inserting
(11) into (8) and (9), while $T$ is obtained with the help of (2). This yields
the following forms

$$n^{(m)}=\frac{n_i^{^{(m)}}}{y^3\,\Gamma},\qquad
n^{(r)}=\frac{n_i^{^{(r)}}}{y^3\,\Gamma} \, , \eqno(17a)$$

$$T= \frac{T_i}{y}\,\Psi \, , \eqno(17b)
$$

$$\rho= \rho^{(m)}+\rho^{(r)}=\left[\frac{\rho_i^{(m)}}{y^3}
+\frac{\rho_i^{(r)}}{y^4}\,\Psi\right]\,\frac{1}{\Gamma}\, , 
\eqno(17c) $$

$$p= \frac{\rho_i^{(r)}}{3y^4} \,\frac{\Psi}{\Gamma} \, ,\eqno(17d)
$$

$$P= \frac{\rho_i^{(r)}}{6y^4}\,\frac{\Phi}{\Gamma} \, , \eqno(17e)$$

\noindent where the functions $\Gamma$, $\Psi$ and $\Phi$ are 
given by

$$\Gamma\equiv \frac{Y'/Y}{Y'_i/Y_i}\qquad \Psi\equiv 1+\frac{\left(1-\Gamma
\right)}{3(1+\Delta_i^{^{(r)}})} ,\qquad
\Phi\equiv
1+\frac{\left(1-4\Gamma
\right)}{3(1+\Delta_i^{^{(r)}})}. \eqno(18)$$

The solutions characterized by (2), (4), (5)-(18) become determinate once
$\Gamma$ above is
obtained in terms of $y$ from (13) for given initial value functions
$\rho_i^{(m)}$, $\rho_i^{(r)}$ (re-expressed in terms of the quantities $
\Delta_i^{^{(m)}},\Delta_i^{^{(r)}},\,\epsilon$).  This transforms
(18) into

$$\Gamma= 1+3A\Delta_i^{^{(m)}}+3B\Delta_i^{^{(r)}} \, , \eqno(19a)$$
$$\Psi= 1-\frac{A\Delta_i^{^{(m)}}
+B\Delta_i^{^{(r)}}}{1+\Delta_i^{^{(r)}}} \, , \eqno(19b)$$
$$\Phi= \frac{-4A\Delta_i^{^{(m)}}
+\left(1-4B\right)\Delta_i^{^{(r)}}}{1+ \Delta_i^{^{(r)}}} \, , 
\eqno(19c)$$

\noindent where

$$A= \frac{1}{3y^2}\left[y^2-4\epsilon
y-8\epsilon^2-\frac{\sqrt{y+\epsilon}}{\sqrt{1+\epsilon}}
\left(1-4\epsilon-8\epsilon^2 \right) \right] \, ,
\eqno(20a)$$

$$B=
\frac{\epsilon}{y^2}\left[y+2\epsilon-\left(1+2\epsilon\right) \frac{\sqrt{y+
\epsilon}}
{\sqrt{1+\epsilon}}\right] \, ,
\eqno(20b)$$

\noindent with $\Delta_i^{(m)},\, \Delta_i^{(r)}$ given by (15).
The kinematic parameters $\sigma,\,\Theta$ follow by inserting
(11) with $y=Y/Y_i$ and (19a) into (6a) and (6b)

$$\sigma=
-\frac{\sqrt{\mu}\sqrt{y+\epsilon}\left[A_{,y}\,\Delta_i^{^{(m)}}+ B_{,y}\,
\Delta_i^{^{(r)}}\right]}{y^2\left[1+3\left(A\Delta_i^{^{(m)}}+
B\Delta_i^{^{(r)}}\right)\right]} \, ,
\eqno(21)$$

$$\frac{\Theta}{3}=
\frac{\sqrt{\mu}\sqrt{y+\epsilon}\left[1+\left(3A+yA_{,y} \right)
\Delta_i^{^{(m)}}+
\left(3B+yB_{,y}\right)\Delta_i^{^{(r)}}\right]}{y^2\left[1
+3\left(A\Delta_i^{(m)}
+B\Delta_i^{(r)}\right) \right]}, \eqno(22)
$$

\noindent where $A_{,y},\,B_{,y}$ are the derivatives of $A,B$ in (20) with
respect to $y$. Given a set of initial conditions specified by
$\epsilon,\,\Delta_i^{^{(m)}},\,\Delta_i^{^{(r)}}$, equations (17) and
(19)-(22) provide fuly determined forms of the state and geometric variables
as functions of $y$ and the chosen initial conditions.

The solutions presented so far contain a FLRW particular case, obtained by
setting in (11) and (12)
$n_i^{(m)}=\bar n_i^{(m)}$, $n_i^{(r)}=\bar n_i^{(r)}$ and $T_i=\bar T_i$,
where $\bar
n_i^{(m)},\,\bar n_i^{(r)},\,\bar T_i$ are arbitrary positive constants.
Under this
parameter specialization, (13) holds with $Y=R(t)f(r)$ (so that $y=R(t)$)
and (5) becomes
a FLRW metric. This leads to

$$\left\langle\rho_i^{(m)}\right\rangle=\bar \rho_i^{(m)},\quad
\left\langle\rho_i^{(r)}\right\rangle=\bar \rho_i^{(r)},\quad
\Gamma=\Psi=1,\quad \Phi=0,$$

\noindent where $\bar \rho_i^{(m)}=mc^2\bar n_i^{(m)}$, $\bar
\rho_i^{(r)}=3\bar
n_i^{(r)}k_{_B}\bar T_i$, so that $T=T(t)$, $\rho=\rho(t)$, $p=p(t)$ and
$P=0$, with (3)
becoming a perfect fluid tensor where $\rho$ and $p$ satisfy (2). The FLRW
limit can also
be characterized by
$\Delta_i^{^{(m)}}=\Delta_i^{^{(r)}}=0$, and so, from (11), equations (21)
and (22) become: $\sigma=0$ and $\Theta/3=\dot y/y=\dot R/R$. Another
limit is that of LTB dust solutions, obtained by setting $T_i=0$ in (12b)
and (11), so that (17) becomes
$T=p=P=0$ and $\rho=mc^2n^{(m)} $.

\section{Density contrasts and regularity conditions.}
Since the radial dependence of all state and geometric variables is
sensitive to
$\Delta_i^{^{(r)}}$ and $\Delta_i^{^{(m)}}$ defined in (15), it is important
to provide an
interpretation for these quantities. From (14) and (15), it is evident
that $\Delta_i^{^{(r)}}$ and $\Delta_i^{^{(m)}}$ are effective ``gauges'' of
the  deviation of $\rho_i^{(m)},\,\rho_i^{(r)}$
from their volume averages for every closed interval in the range of the
integration variable $r$ along the initial hypersurface $t=t_i$.  The signs of
these quantities characterize initial density profiles, with ``density lumps''
as these densities decrease ($[\rho_i^{(m)}]'<0,\,[\rho_i^{(r)}]'<0$) or
``density voids'' as they increase ($[\rho_i^{(m)}]'>0,\,[\rho_i^{(r)}]'>0$).
Also, with the help of Rolle's theorem applied to (14) we find that
$\Delta_i^{^{(r)}}$ and $\Delta_i^{^{(m)}}$,
as adimensional functions of $r$, are constrained by the maximal density
contrasts in terms of

$$ |\Delta_i^{^{(m)}}| \le
\frac{\rho_i^{(m)}{}^{\hbox{max}}}{\rho_i^{(m)}{}^{\hbox{min}}}-1 \, ,$$
$$ |\Delta_i^{^{(r)}}| \le \frac{\rho_i^{(r)}{}^
{\hbox{max}}}{\rho_i^{(r)}{}^{\hbox{min}}}- 1 \, ,$$

\noindent where the superindices ``max'' and ``min'' respectively indicate
the maximal and minimal values of $\rho_i^{(m)},\,\rho_i^{(r)}$ in any
interval $0\le r$ along the hypersurface $t=t_i$. Small initial density
contrasts obviously imply

$$\rho_i^{(m)}{}^{\hbox{max}}\approx
\rho_i^{(m)}{}^{\hbox{min}},\qquad \rho_i^{(m)}\approx \left\langle
\rho_i^{(m)}
\right\rangle\quad\Rightarrow\quad |\Delta_i^{^{(m)}}|\ll 1, \eqno(23a)$$
$$\rho_i^{(r)}{}^{\hbox{max}}\approx
\rho_i^{(r)}{}^{\hbox{min}},\qquad \rho_i^{(r)}\approx \left\langle
\rho_i^{(r)}
\right\rangle\quad\Rightarrow\quad |\Delta_i^{^{(r)}}|\ll 1, \eqno(23b)$$

$$ \mu\approx \frac{\kappa}{3}\,\rho_i^{(m)}, \qquad \epsilon\approx
\frac{\rho_i^{(r)}}{\rho_i^{(m)}}\approx
\frac{3n_i^{(r)}}{n_i^{(m)}}\,\frac{k_{_{B}}T_i}{mc^2}\approx
10^{-9}\,\frac{k_{_{B}}T_i}{mc^2}, \eqno(23c)$$

\noindent allowing us to consider a formal analogy between
$\Delta_i^{^{(r)}}$  and $\Delta_i^{^{(m)}}$ and energy density ``exact''
initial perturbations. This is further
reinforced from the definitions in (15), and by remarking that the FLRW
``background'' follows by ``turning the perturbations off'', that is, setting:
$\Delta_i^{^{(m)}}=\Delta_i^{^{(r)}}=0$.

An important restriction that the solutions must satisfy is the following
regularity condition

$$\Gamma\equiv \frac{Y'/Y}{Y'_i/Y_i}>0,  \eqno(24)$$

\noindent which prevents negative densities $n^{(r)},\,n^{(m)}$, as well as
the occurrence of a shell crossing singularity \cite{gw}. This singularity is
characterized by unphysical
behavior because $\Gamma$ appears in the denominator  of equations (17a),
(17c), (17d) and (17e),
but does not appear in (17b). Therefore, if $\Gamma= 0$, the densities,
pressure and viscous
pressure diverge with $T$ finite (in general), a totally unacceptable
situation that can be avoided by considering only the range of evolution of
the models to spacetime seccions with $t\ge t_i$ satisfying (24).  The
fulfilment of (24) depends on the functions $A,B$ and on the magnitudes of the
initial density contrasts gauged by $\Delta_i^{^{(r)}}$ and
$\Delta_i^{^{(m)}}$. This will be examined further
ahead together with the conditions for thermodynamical consistency.

\section{Thermodynamical consistency}

The models derived and presented in the previous sections must be
compatible with a
suitable thermodynamical formalism. For this purpose, it is advisable to
leave aside the
``conventional" theory of irreversible thermodynamics \cite{eck},
\cite{ll}, whose transport equations are unphysical as they violate
relativistic causality of
the dissipative signals as well as stability of the equilibrium states 
(seee.g. \cite{hl1}, \cite{hl2} and \cite{hl3}). We shall 
consider instead ``Extended Irreversible Thermodynamics'' (EIT)
\cite{wi},\cite{wjs},\cite{ddj},\cite{jcl}, a theory free of such serious
drawbacks \cite{hl3},
and so a more adequate theoretical framework for the models. According to
EIT, the entropy
of a system away from thermodynamical equilibrium depends not only on the
``conserved"
(equilibrium) variables (i.e. particle number densities, energy density and
so on), but also
into the non-equilibrium fluxes (i.e. heat flux, and bulk and shear
dissipative stresses).
This theory is supported by Kinetic Theory of gases, Information Theory and
by the
Theory of Hydrodynamical Fluctuation - see
\cite{jcl} for a detailed description. When shear viscosity is the only
dissipative agent, the corresponding generalized entropy $ \; s \,$ of
radiation plus matter
obeying the usual balance law with non-negative divergence, up to second
order in $\, \Pi^{ab}
\,$ takes the form

$$s= s^{(e)}+\frac{\alpha}{nT}\,\Pi_{ab}\Pi^{ab},\quad \Rightarrow \quad
(snu^a)_{;a}\geq 0,
\eqno(25)$$

\noindent where $\,s^{(e)}\,$ is obtained from the integration of the
equilibrium Gibbs
equation, $n=n^{(r)}+n^{(m)}$ and $\alpha$ is a phenomenological
coefficient to be specified
later. The evolution of the viscous pressure is, in turn, governed
by the transport
equation

$$\tau \dot\Pi_{cd}\,h^c_ah^d_b+\Pi_{ab}\left[
1+{1\over2}T\eta\left({{\tau}\over{T\eta}}  \,
u^c\right)_{;c}\right]+ 2\eta\,\sigma_{ab}= 0 \, , \eqno(26) $$

\noindent
where $\, \eta \,$ and $ \, \tau$ are the coefficient of shear viscosity
and the relaxation
time of shear viscosity, respectively. The former as well as other related
quantities can be
obtained by a variety of means \cite{wei} including Kinetic Theory,
Statistical Mechanics or
both  \cite{uwi},\cite{dd1},\cite{dd2},\cite{zim}. The relaxation time, $\,
\tau \,$, is
related to and larger than the mean collision time between particles and it
may, in
principle, be estimated by collision integrals provided the interaction
potential is known.
As a physical reference to infere the form these coefficients might take,
consider the
``radiative gas'', with $p, \,
\rho$ satisfying (1) (or the approximation (2)). For the radiative gas the
forms of $\eta, \,
\alpha$ in terms of the relaxation time of the dissipative process $\tau$ are

$$\eta_{_{(rg)}}= {4\over{5}}p^{(r)}\tau,\qquad \alpha_{_{(rg)}}= -
{\tau\over{2\eta_{_{(rg)}}}}= -\frac{5}{8p^{(r)}} \, \eqno(27)$$

\noindent
where $p^{(r)}$ is either $n^{(r)}k_{_B}T$ or $a_{B}T^4$ and the subscript
``$(rg)$''
emphasizes that these quantities are specific to the radiative gas.

We verify now the compatibility of (13)-(22) with (25)-(27). Integrating
the equilibrium
Gibbs equation and substituting (27) into (25), we obtain

$$s=
\frac{4a_{_B}T_i^3}{3n_i^{(r)}}+k_{_B}\ln\left[\frac{n^{(r)}_i}{n^{(r)}}
\left(\frac{T}{T_i}
\right)^3\right]-
\frac{15k}{8}
\left(\frac{P}{p}\right)^2 $$
$$=
\frac{4a_{_B}T_i^3}{3n_i^{(r)}}+k_{_B}\ln\left[\frac{\Psi^3}{\Gamma}\right]
-\frac{15k}{32}
\left(\frac{\Phi} {\Psi} \right)^2 \, , \eqno(28)$$

\noindent where the approximation: $n= n^{^{(m)}}+n^{^{(r)}}\approx
n^{^{(r)}}$ was used and
the initial value of $s^{(e)}$ has been set to be the equilibrium entropy
per photon. Equation
(28) reflects the fact that we are neglecting the contribution of the
entropy due to
non-relativistic particles, a justified approximation since the latter are
much less abundant
than the photons.

Ideally, the transport equation (26) should be satisfied for $\eta$ having
the form (27),
associated to the radiative gas, and the relaxation time, $\tau$, given by
collision
integrals obtained from Kinetic Theory. However, as mentioned in the
introduction, and in
order to obtain exact expressions for all thermodynamical parameters, we
will assume $\eta$
given by (27) and deduce $\tau$ from the fulfilment of (26). This yields

$$\tau =\frac{-\Psi\Phi}{\sigma}\; \frac{{\textstyle{{9} \over
{4}}}\left[1+\Delta_i^{^{(r)}}\right]^2}{{\textstyle{{4} \over
{5}}}\left[3+4\Delta_i^{(r)}+{\textstyle{{13} \over
{32}}}\Gamma\right]^2+{\textstyle{{171} \over
{256}}}\Gamma^2 }. \eqno(29)$$

\noindent While there is no need to justify $\eta$ given by (27), this form
of $\tau$ is
acceptable as long as (29) satisfies the requirement of a relaxation
parameter: it must be a
positive quantity and must comply with a positive entropy production law
$\dot s\geq 0$. It is desirable that $\tau$ should somehow relate or approach
its definition as a collision
integral and that its behavior be qualitatively analogous to a suitable
mean collision time, therefore it should be an increasing (decreasing)
function if the fluid is expanding (collapsing). Evaluating
$\dot s$ from (28) and comparing with (29), we obtain the following
relation between $\dot s$ and $\tau$

$$\dot s= \frac{15 k_{_B}}{4\tau}\left(\frac{P}{p}\right)^2= \frac{15
k_{_B}}{16\tau}
\left(\frac{\Phi} {\Psi}\right)^2 , \eqno(30)$$

\noindent consistent with the general relation \cite{ddj} $(nsu^a)_{;a}=
\Pi_{ab}\Pi^{ab}
/(2\eta nT) $ associated with (25) and (26). As a consequence of (30),
$\dot s>0$ and
$\tau>0$ imply each other. Also, from the form of (29), necessary
and sufficient conditions for positive $\tau,\,\dot s,\,p,\,T$ can be given by

$$\Psi>0 , \eqno(31a)$$
$$\sigma\Phi<0 , \eqno(31b)$$

\noindent while the condition ensuring concavity and stability of $s$ can
be phrased for an
expanding fluid configuration as the requirement that $\dot s$ decreases
for increasing $\tau$
($\dot\tau>0 \Leftrightarrow \ddot s<0$). From (30)-(31), this follows as

$$\dot \tau>0,\qquad {\ddot s\over\dot
s}= {{2\sigma\Gamma}\over{3\Psi\Phi}}\,{{\left\langle
\rho_i^{(r)}\right\rangle}\over
{\rho_i^{(r)}}}\left[1+{{\left\langle \rho_i^{(r)}\right\rangle}\over
{3\rho_i^{(r)}}}\right]-{\dot\tau\over{\tau}}<0. \eqno(31c)$$

\noindent So that, if (24), (31a) and (31b) hold, then (31c) reduces to
$\dot \tau>0$.

Since $\tau$ is a thermodynamic relaxation parameter, it is important to
compare it with
another natural timescale of the models: the Hubble expansion time defined by

 $$t_{_H}=\frac{3}{\Theta}. \eqno(32a)$$

\noindent where $\Theta$ follows from (22). Such comparison  should provide an
insight into the timescales associated with the mixture interaction and
decoupling. However, strictly speaking, the
criterion for interaction and decoupling in cosmological gas mixtures is
not given by
comparing $\tau$ and $t_{_H}$ but by comparing the latter with the timescales
associated with the various reaction rates of the radiative processes involved,
particularly the photon mean collision time $t_{_\gamma}$ obtained from Thomson
scattering  \cite{kotu},\cite{pad}. Hence, we can consider this
relaxation time as approximately gauging the interactivity of the matter
mixture by
demanding that for a range of the evolution of the mixture, approximating
its interactive
range, we must have

$$\tau <  t_{_H}\eqno(32b)$$

\noindent
while, as the mixture evolves and the components decouple, eventually

$$\tau >  t_{_H}. \eqno(32c)$$

\noindent
Since  $\tau$ in (27) and (29) must behave qualitatively similar to
$t_{_\gamma}$\cite{schweitzer}, the comparison in (32b-c) should yield
qualitatively analogous results as a similar comparison between $t_{_H}$ and
$t_{_\gamma}$. The temperature associated with the passage from (32b) to (32c)
(obtained from (17b)) should approximate the decoupling temperature obtained
by the condition $t_{_\gamma} =t_{_H}$. This point is examined in section XI.

Equations (31)-(32), together with the regularity condition (24),  provide the
necessary and sufficient conditions for a theoretically consistent
thermodynamical description
of the solutions within the framework of EIT and Kinetic Theory applied to
the radiative gas. We examine
the effect of these
conditions in the following section.

\section{Thermodynamically consistent models}

\subsection{Conditions (24) and (31a)}

From (19a) and (19b), the fulfilment of (24) and (31a) is equivalent
to the following condition

$$ -\frac{1}{3}< A\Delta_i^{(m)}+B\Delta_i^{(r)} < 1+\Delta_i^{^{(r)}}.
\eqno(33)$$

\noindent From (20), since the functions $A$ and $B$ diverge as $y\to 0$
there are necessarily values of $y$ for which (33) is violated. However, from
(19a-b) we have: $\Gamma_i=\Psi_i=1$, and the range of $y$ we are interested
is $y\ge 1$. As shown by figures 1a and 1b, displaying the implicit plots of
$\Gamma=0$ and $\Psi=0$ for an initial
temperature $T_i\approx 10^6$ K (from (23c), $\epsilon\approx 10^3$), there
are no zeros of $\Gamma$ and $\Psi$ for $y\ge 1$, $|\Delta_i^{^{(m)}}|\le 1$
and $|\Delta_i^{^{(r)}}|\le 1$. Therefore, (33) holds in the range of
interest for a large class of initial conditions.

\subsection{Conditions (31b), (31c) and (32)}

For the examination of these conditions we will assume that (33) holds for
$y\ge 1$,\, $|\Delta_i^{^{(m)}}|\le 1$ and $|\Delta_i^{^{(r)}}|\le 1$ (see
figures 1a and 1b). Then, with the help of (19)-(22), condition (31b) is
equivalent to

$$C\equiv
4AA_{,y}\left[\Delta_i^{^{(m)}}\right]^2+
(4B-1)B_{,y}\left[\Delta_i^{^{(r)}}\right]^2 +
\left[4AB_{,y}+(4B-1)A_{,y}\right]\Delta_i^{^{(m)}}\Delta_i^{^{(r)}}<0.
\eqno(34a)$$

\noindent An insight into this expression follows by looking at its initial
value

$$C_i=\frac{-\Delta_i^{^{(r)}}}{2(1+\epsilon)}\,\left[\Delta_i^{^{(r)}}+\epsilon
\Delta_i^{^{(m)}} \right]\, ,
\eqno(34b)$$
\noindent and its asymptotical behavior as
$y\gg 1$

$$C\approx\frac{4\epsilon\left[\Delta_i^{(s)}\right]^2}{9y^2}+
\frac{4\lambda\Delta_i^{(s)}}
{\sqrt{1+\epsilon}y^{5/2}}+\frac{35}{9}
\frac{\epsilon\lambda\Delta_i^{(s)}}{\sqrt{1+\epsilon}y^{7/2}}$$
$$-\frac{
\lambda_1\left[\Delta_i^{(s)}\right]^2+
24\lambda_2\Delta_i^{(r)}\Delta_i^{(s)}+
9\left[\Delta_i^{^{(r)}}\right]^2}{24(1+\epsilon)
\,y^4}$$
$$+\frac{64}
{63}\frac{\epsilon^2\lambda\Delta_i^{(s)}}{\sqrt{1+\epsilon}y^{9/2}}-\frac{
\epsilon}{18}
\frac{
\lambda_1\left[\Delta_i^{(s)}\right]^2+
24\lambda_2\Delta_i^{(r)}\Delta^{(s)}+
9\left[\Delta_i^{^{(r)}}\right]^2}{(1+\epsilon)\, y^4}$$
$$-\frac{55}{8}
\frac{\epsilon^3\lambda\Delta_i^{(s)}}{\sqrt{1+\epsilon}y^{11/2}}+
{\cal O}(y^{-6}), \eqno(34c)$$

\noindent where
$$\Delta_i^{(s)}\equiv {\textstyle{{3} \over
{4}}}\Delta_i^{^{(r)}}-\Delta_i^{^{(m)}}, \eqno(35)$$
\noindent and
$$\lambda\equiv
\Delta_i^{^{(m)}}\lambda_2-  3\epsilon(1+2\epsilon)\Delta_i^{^{(r)}}, $$
$$\lambda_1\equiv 1-8\epsilon+2\epsilon^3+2\epsilon^4 , 
\qquad \lambda_2 \equiv 8\epsilon^2+4\epsilon-1. $$

\noindent As shown by (34c), the quantity $\Delta_i^{(s)}$ defined in (35)
plays a fundamental role with regards to the fulfilment of (31b) and (31c).
This suggest classifying initial conditions in terms of $\Delta_i^{(s)}$. We
shall examine (31b), (31c) and (32) for
each case $\Delta_i^{(s)}=0$ and $\Delta_i^{(s)}\ne 0$ separately.

\subsubsection{The case $\Delta_i^{(s)}=0$}

>From (34c), the condition $\Delta_i^{(s)}=0$ is necessary and sufficient for
having $C<0$ asymptotically. In fact, substituting $\Delta_i^{(s)}=0$ into
(34a), leads to the following dramatic simplification  of $C$

$$C= -\frac{\left[\Delta_i^{^{(r)}}\right]^2\left(3y+4\epsilon\right)}{8(1+
\epsilon)\,y^5} \, ,
\eqno(36)$$

\noindent so that $\Delta_i^{(s)}=0$ is a sufficient condition for the
fulfilment of (24), (31a) and (31b) along the range $y\ge 1$, for
$|\Delta_i^{^{(r)}}|\le 1$ and $|\Delta_i^{^{(m)}}|<1$.

Condition (31c) is also satisfied, since $\tau$ increases monotonously
along the fluid worldlines, behaving asymptotically as: $\tau\approx y^{3/2}$.
Hence, those models whose initial conditions satisfy $\Delta_i^{(s)}=0$ can be
characterized as the subclass of  models
complying with the conditions of thermodynamical consistency in the
asymptotic range of $y$. This point is consistent with the fact that
$\Delta_i^{(s)}=0$ is a necessary and
sufficient condition for having $|P|/p=|\Phi|/(2\Psi)\to 0$ and $\dot s\to 0$
as $y\to\infty$, so that the fluid layers evolve towards an asymptotic
equilibrium state.
However, using $t_{_{H}}=3/\Theta$ calculated by inserting $\Delta^{(s)}=0$
into (22),
we have $\tau/t_{_H}<1$ for all the evolution of the fluid, thus failing to
comply with
(32b-c). This is illustrated by figure 2, and implies that the relaxation
time $\tau$
cannot be associated with a radiation matter mixture whose components
interact and then
decouple.

\subsubsection{The case $\Delta^{(s)}\ne 0$}

If $\Delta^{(s)}\ne 0$, condition (31b) cannot hold along the full range of
$y$ (because of
(34b)), but might hold along a restricted range of physical interest $1\le
y \le y_*$ for
which the mixture could be in the interactive stage. This situation is not
incompatible
with the thermodynamical arguments of the previous section, since the
phenomenology of the
radiative gas model strictly applies if the mixture components interact.
The fact that
condition (31b) can hold for $1\le y \le y_*$  follows from evaluating the
sign of $C_i$
given by (34a), a quantity that can be negative (so that $\tau_i>0$) for a
wide range of
acceptable situations  (for example, if $\Delta_i^{^{(r)}}$ and
$\Delta_i^{^{(m)}}$ have the same sign). However, as $y$ increases, $\tau$
either changes sign or
diverges positively, depending on the zeroes of
$\Phi$ and $\sigma$. Since we are assuming that (24) and (31a) hold, the
sign of
$\tau$, as given by (29),  depends only on the quotient $\Phi/\sigma$, and
so the behavior of
$\tau$ is strongly related to the signs and zeroes of these functions. If
$\tau_i>0$ and, as
$y$ increases, there is a zero of $\Phi$ for $\sigma\ne 0$, then $\tau$
passes from positive
to negative, but if the zero of $\sigma$ appears first, then $\tau$
diverges positively. A
zero of $\Phi$ (with $\sigma\ne 0$) might be compatible with (31b)
($\tau>0$), but violates
(31c) and so is unacceptable.  However, a zero of $\sigma$ (with $\Phi\ne
0$) is acceptable,
since $\tau$ would diverge positively (as $y\to y_*$) and so would be a
positive and
increasing function along the range $1\le y \le y_*$. In order to verify if
this type of
evolution is possible, it is necessary to gather information on the zeros
of $\Phi$ and
$\sigma$. Since these expressions are cumbersome, it is convenient to
examine their zeros
graphically, and so we have plotted in figures 3a and 3b the solutions of
the implicit
equations
$\Phi=0$ and $\sigma=0$, in terms of $\Delta_i^{^{(m)}}$,
$\Delta_i^{^{(r)}}$ and
$\log_{10}(y)$,
while figure 4 displays the sectors in the plane
$\Delta_i^{^{(m)}},\,\Delta_i^{^{(r)}}$ where
$\Phi=0$ and $\sigma=0$ occur. As these figures reveal, sufficient
conditions for
the desired evolution are given by

$$\Delta_i^{(s)}>0\qquad \hbox{for}\quad \Delta_i^{^{(r)}}>0 \, \eqno(37a)$$
$$\Delta_i^{(s)}<0\qquad \hbox{for}\quad \Delta_i^{^{(r)}}<0. \eqno(37b)$$

\noindent Therefore, if $\Delta_i^{(s)}\ne 0$, conditions (37) are sufficient
for the fulfillment of (31b) and (31c) along the range 
$1\le y\le y_*$,  where $y_*$ is a
zero of $\sigma$. Conditions (32) are satisfied under the restrictions
(37). This will be
discussed further ahead in section X.

\section{Initial conditions and exact initial perturbations.}

The behavior of the quantity $\Delta_i^{(s)}$ given by (35) determines the
set of initial
conditions that characterize the thermodynamical consistency of the models.
Since this
quantity plays such an important role, it must be related to a physically
significant
property along the initial hypersurface $t=t_i$. From (35), using (12), (14)
and (15), we obtain

$$\Delta_i^{(s)}=\frac{3}{4}\Delta_i^{^{(r)}}-\Delta_i^{^{(m)}}=
\frac{1}{4}+\frac{\left[\log(W^{3/4}/M)\right]{}'}{\left[\log(Y_i^3)\right]{}'}
=\frac{d\left[\log\Sigma_i\right]}{d\left[\log Y_i^3\right]},
\eqno(38a)$$
\noindent where $$\Sigma_i\equiv \frac{\left\langle \rho_i^{(r)}
\right\rangle^{3/4}}{\left\langle \rho_i^{(m)}
\right\rangle}. \eqno(38b)$$

\noindent
In order to provide an interpretation for (38), we remark (from (17), (19)
and (28)) that the
entropy per photon along the initial hypersurface

$$s_i=\frac{4a_{_B}T_i^3}{3n_0^{(r)}}-\frac{15}{32}k_{_B}\left[\frac{\Phi_i}
{\Psi_i}\right]^2
=\frac{4a_{_B}T_i^3}{3n_i^{(r)}}-
\frac{15}{32}k_{_B}\left[\frac{\Delta_i^{^{(r)}}}
{1+\Delta_i^{^{(r)}}}\right]^2
$$

\noindent is very close to its equilibrium value, since its off equilibrium
correction is
proportional to $[\Delta_i^{^{(r)}}]^2$, a very small quantity since
$\Delta_i^{^{(r)}}$ is already
assumed to be small. Therefore, using the approximations associated with
small density
contrasts (equations (23a-c)), we obtain

$$\rho_i^{(r)}\approx 3a_{_B}T_i^4,\qquad \Sigma_0 \propto
\frac{\left\langle a_{_B}T_i^4
\right\rangle^{3/4}}{mc^2\left\langle n_i^{(m)} \right\rangle}\propto
\frac{\left\langle s_i
\right\rangle}{\left\langle n_i^{(m)}
\right\rangle} \approx \left\langle \frac{s_i}{n_i^{(m)}}
\right\rangle \, , \eqno(39)$$

\noindent implying that $\Sigma_i$ is proportional to the ratio of the
averages of
photon entropy and baryon number density, which (for small density
contrasts) is roughly
equivalent to the averaged ratio of these quantities. Hence, condition
$\Delta_i^{(s)}=0$
roughly means an initial hypersurface with constant averages of photon
entropy per baryon, while condition (37) roughly means (see equations (20))
that the
sign of the spatial gradient of the average of photon entropy per baryon
must agree
with the sign of the gradient of the initial photon energy density: as $r$
increases along
$t=t_i$, it must increase for a density void ($[\rho_i^{(r)}]'>0$) and
decrease for a density
lump ($[\rho_i^{(r)}]'<0$).

The assumption of small density contrasts leads to a natural comparison
with the Theory of
Perturbations of a FLRW background, in the isochronous gauge and considering a
mixture of radiation and non-relativistic matter, see
\cite{bor},\cite{kotu},\cite{pad},\cite{peeb},\cite{kosa},\cite{efsta}. Under
these assumptions,
matter and radiation densities are given by

$$\rho^{(m)}=\bar\rho^{(m)}\left[1+\delta^{(m)}\right], \eqno(40a)$$
$$\rho^{(r)}=\bar\rho^{(r)}\left[1+\delta^{(r)}\right], \eqno(40b)$$

\noindent where $\bar\rho^{(m)},\,\bar\rho^{(r)}$ are the respective
densities in a FLRW
background and $\delta^{(m)}\ll 1,\,\delta^{(r)}\ll 1$ are the
perturbations. The gauge
invariant quantity

$$\delta^{(s)}=\frac{3}{4}\delta^{(r)}-\delta^{(m)} , \eqno(41a)$$

\noindent which formally resembles (35), defines the fluctuations of photon
entropy per
baryon and leads to the classification of perturbations as

$$\hbox{``Adiabatic''}\qquad\qquad \delta^{(s)}=0 ,  \eqno(41b)$$
$$\hbox{``Isocurvature''}\qquad\qquad
\delta^{(s)}=\delta^{(m)}\left[1+\frac{3\rho^{(m)}}{4\rho^{(r)}}\right].
\eqno(41c)$$

\noindent In order to establish a comparison with Perturbation Theory, consider
defining the initial densities along the lines of (40)

$$\rho_i^{(m)}=\bar\rho_i^{(m)}\left[1+\delta_i^{(m)}\right]
\quad\Rightarrow\quad \left\langle\rho_i^{(m)} \right\rangle
=\bar\rho_i^{(m)}\left[1+\left\langle\delta_i^{(m)}\right\rangle\right],
\eqno(42a)$$
$$\rho_i^{(r)}=\bar\rho_i^{(r)}\left[1+\delta_i^{(r)}\right]
\quad\Rightarrow\quad \left\langle\rho_i^{(r)} \right\rangle
=\bar\rho_i^{(r)}\left[1+\left\langle\delta_i^{(r)}\right\rangle\right],
\eqno(42b)$$

\noindent where $\bar\rho_i^{(m)},\,\bar\rho_i^{(r)}$ are now the constant
values of the initial densities in a FLRW background and 
$\mid \delta_i^{(m)}\mid \ll 1,\,\mid \delta_i^{(r)}\mid \ll 1$ 
are exact initial perturbations. Inserting (42) into (38) leads
to

$$\Sigma_i\approx
\frac{\bar s_i}{\bar n_i^{(m)}}
\frac{1+\textstyle{{3} \over
{4}}\left\langle\delta_i^{(r)}\right\rangle}{1+\left\langle\delta_i^{(m)}
\right\rangle}
\quad\Rightarrow\quad\frac{\Delta_i^{(s)}}{Y_i^3}\approx\frac{d}{dY_i^3}
\left\langle
\frac{3}{4}\delta_i^{(r)}-\delta_i^{(m)}
\right\rangle\approx
\frac{d}{dY_i^3}\left\langle\delta_i^{(s)}\right\rangle  , \eqno(42c)$$

\noindent where $\bar s_i\propto [\bar\rho_i^{(r)}]^{3/4}$ and $\bar
n_i^{(r)}$ are the
photon entropy and baryon number density of the FLRW background in the
hypersurface $t=t_i$.
The rhs of (42c) illustrates that, under the assumptions (42a-b) of a
perturbative treatment,
$\Delta_i^{(s)}$ approximately reduces to the radial gradient of
$\delta^{(s)}$ evaluated in
$t=t_i$. However, the comparison with perturbations must be handled with
caution, since in
the case of (35) and (36)-(39) we are dealing with quantities and relations
strictly
defined in the initial hypersurface, and so determining initial conditions
in terms of
averages of initial value functions. The Theory of Perturbations, on the
other hand,
traces the evolution of $\delta^{(m)},\,\delta^{(r)}$ for all $t$. In
particular, we must
be specially careful if we borrow the terminology of perturbations in (41)
to characterize
the cases $\Delta_i^{(s)}=0$ and $\Delta_i^{(s)}\ne 0$, since in our case a
relation like
$\Delta_i^{(s)}=0$ specifies initial conditions and will not be satisfied (in
general) for
$t>t_i$. Another delicate point refers to the current use of the term
``adiabatic'' for
the case (41b), meaning perturbations that conserve photon entropy. Strictly
speaking, these perturbations should be denoted as ``isentropic'' or
``reversible'', since an
adiabatic process need not be isentropic (or reversible). It is quite
possible that this
conceptual vagueness follows from the fact that most papers in Perturbation
Theory, either
assume thermal equilibrium, or have incorporated dissipative processes
without the necessary
rigour. Hopefully, recent work \cite{matr} on these lines might be
helpful to clarify these
issues. Raising this point is relevant because the models presented in this
paper assume a
fluid evolving along adiabatic but irreversible (non-isentropic) processes,
therefore initial
conditions like
$\Delta_i^{(s)}=0$ (formally analogous to (41b)) do not imply a conserved
photon entropy,
but a roughly constant average of photon entropy at $t=t_i$, but not (in
general) for
$t>t_i$. Never the less, for the sake of mantaining continuity with
currently established
terminology, we shall use the formal analogy between (41) and (42) in the
examination of the cases $\Delta_i^{(s)}=0$ and $\Delta_i^{(s)}\ne 0$.
Developing further the comparison with perturbations on a FLRW background is
relevant and interesting. However, such a task requires a comprehensive and
detailed elaboration and so will be carried on elsewhere.

\section{The case $\Delta_i^{(s)}\ne 0$, an interactive mixture.}

From section VIII we know that if (37a-b) hold, then $\tau$ diverges
positively as $y\to y_*$
where $y_*$ is a zero of $\sigma$. We still
need to know
if $\tau/t_{_H}<1$ along the range $1\le y<\tilde y$, where the set of values
$y=\tilde y<y_*$ are characterized by $\tau/t_{_H}=1$, approximately marking
the decoupling of matter and radiation. A
sufficient condition for this type of evolution follows from evaluating
$[\tau/t_{_H}]_i$ at the initial hypersurface

$$\left[ {{\tau  \over {t_{_H}}}} \right]_i={{5\Delta_i^{(r)}\left [
{1+\Delta_i^{(r)}}
\right]\left[ {(3+4\varepsilon )\Delta_i^{(r)}-4\Delta_i^{(s)}+8\left(
{1+\varepsilon }
\right)} \right]} \over {\left[ {(3+4\varepsilon
)\Delta_i^{(r)}-4\Delta_i^{(s)}(3+4)}
\right]\left[ {36+47\Delta_i^{(r)}+16\left( {\Delta_i^{(s)}} \right)^2}
\right]}}\approx {10(1+\epsilon)
\over {9(3+4\epsilon)}}\approx\frac{5}{18}<1 , $$

\noindent where we have eliminated $\Delta_i^{^{(m)}}$ from (35) and assumed
$\Delta_i^{(s)}\ll 1$, $\Delta_i^{^{(r)}}\ll 1$ and $\epsilon\approx
10^{-3}T_i\approx 10^3$. Since $\tau$ diverges as
$y\to y_*$ but $\Theta$ remains finite in this limit, the ratio
$\tau/t_{_H}$, initially smaller, necessarily becomes larger than unity for
$1<y<\tilde y<y_*$.

The currently accepted value of matter and radiation decoupling is
$T_{_D}\approx 4\times 10^3$ K. Assuming $T_i\approx 10^6$ K, the set of
values $y=y_{_D}$ associated with this temperature follow from (17b) by
solving for $y=y_{_D}$ the equation

$$T_{_D}\approx 4\times 10^3
=\frac{10^6}{y_{_D}}\Psi(y_{_D},\Delta_i^{(s)},\Delta_i^{^{(r)}}) ,
\eqno(43)$$

\noindent where $\Psi$ is given by (19b). Assuming small density contrasts,
figure 5
illustrates that this value of $T_{_D}$ is closely associated with
$y_{_D}\approx 10^{2.4}$.
Also, it is evident from figure 3b, that having a zero of $\sigma$ for
values $10^2<y_*< 10^3$ requires a very small deviation from
$\Delta_i^{(s)}=0$. This is illustrated by the
approximated sketch of the level curve $y=y_{_D}=10^{2.4}$ that appears in
figure 4, implying that $\tilde y,\,y_*$ and $y_{_D}$ lie in a a very narrow
sector of the plane $\Delta_i^{^{(m)}},\,\Delta_i^{^{(r)}}$, very close to the
line $\Delta_i^{(s)}=0$. Therefore, for
these values we must have $y_*\approx\tilde y\approx y_{_D}$. However, it
is not clear from figures 3b and 4 how small the deviation from
$\Delta_i^{(s)}=0$ should be. Hence, we have ploted in
figure 6 the implicit equation $\sigma=0$, showing a more precise relation
between the orders of magnitude that the occurrence of the zero of $\sigma$
for $y_{_D}\approx y_*\approx 10^{2.4}$ implies for
$|\Delta_i^{^{(r)}}|,\,|\Delta_i^{(s)}|$.

Another constraint the models have to comply with is the observational
bounds\cite{kotu},\cite{pad} on the anisotropy of the cosmic microwave
radiation (CMR), given by the maximal photon temperature contrast 
$[\delta T/T]_{_D}\approx10^{-5}$. For $y\approx 10^{2.4}$ and
$T_i\approx 10^6$, so that $\epsilon\approx 10^3$, we have: $A\approx
1.1\times 10^{-1}$ and $B\approx 2.2\times 10^{-1}$, were $A$ and $B$ are
given by (20). Also, from (17b) and (19b),
the deviation from the equilibrium FLRW form $T_{eq}=T_i/y$ at $y=y_{_D}$ is
given by

$$\left[\frac{\delta
T_{eq}}{T_{eq}}\right]_{_D}=\left[\frac{T-T_{eq}}{T_{eq}}\right]_{_D}=
\left[\Psi-1\right]_{_D}\approx -\frac{1.1\times
10^{-1}\Delta_i^{^{(m)}}+2.2\times
10^{-1}\Delta_i^{^{(r)}}}{1+\Delta_i^{^{(r)}}}, \eqno(44)$$

\noindent indicating that compliance with the maximal observed temperature
contrast of the CMR constrains the maximal values of
$\Delta_i^{^{(m)}},\,\Delta_i^{^{(r)}}$ to about $10^{-4}$.
Therefore, from figure 6, the corresponding variation range of
$|\Delta_i^{(s)}|$ is $|\Delta_i^{(s)}|<10^{-8}$, and so compatibility with
acceptable values of $T_{_D}$ and the CMR anisotropy implies
$|\Delta_i^{(s)}|\approx |\Delta_i^{^{(r)}}|^2\ll |\Delta_i^{^{(r)}}|$.

Under the analogy between $\Delta_i^{^{(m)}},\, \Delta_i^{^{(r)}}$ with
perturbations in a FLRW background, the case $\Delta_i^{(s)}\ne 0$ seems to
correspond to isocurvature initial
perturbations. However, since $\rho^{(m)}\ll \rho^{(r)}$ before decoupling,
the latter are characterized (see equation (41c)) by
$\delta^{(s)}\approx\delta^{(m)}$ and
$\delta^{(r)}\ll\delta^{(s)}$, and so, following the analogy between (41) and
(42) would disqualify $\Delta_i^{(s)}\ne 0$ as comparable to an isocurvature
initial perturbation. In fact, the
maximal bounds on $\Delta_i^{^{(r)}}$ and $\Delta_i^{(s)}$ obtained in the
previous parragraph yield exactly the opposite behavior. Borrowing the
terminology of Perturbation Theory, the analogy
between (41) and (42) would characterize ``isocurvature'' initial conditions by
$\Delta_i^{^{(r)}}=0$. From figures 3 and 4, this condition is not incompatible
with (31) and (32), but yields a decoupling surface $y_{_D}\approx 10$ with a
decoupling temperature much larger than the accepted value. Hence, the
acceptable values of the case $\Delta_i^{(s)}\ne 0$ cannot be associated with
this type of initial perturbations, but since perturbations are in general
combinations of adiabatic and isocurvature components and we
have $\Delta_i^{(s)}\ll \Delta_i^{^{(r)}}$ and $\Delta_i^{(s)}\ll
\Delta_i^{^{(m)}}$, a more accurate analogy for
$\Delta_i^{^{(m)}},\,\Delta_i^{^{(r)}}$ is that of ``quasi adiabatic''
initial perturbations. Following this analogy, and bearing in mind that
$\Delta_i^{^{(m)}}$ and $\Delta_i^{^{(r)}}$ are defined for an initial
hypersurface characterized by $T_i\approx 10^6$ K.
(still outside the horizon for perturbations of all wave numbers), the
bounds on the magnitude of $\Delta_i^{(s)}$ can be related to bounds on the
amplitude of ``nearly adiabatic'' fluctuations of photon entropy per baryon
generated in the earlier, more primordial, inflationary era
\cite{bor},\cite{kotu},\cite{pad},\cite{efsta},\cite{linde}. Primordial entropy
fluctuations have been examined in connection with the creation of axions in
an inflationary scenario\cite{kotu},\cite{efsta}. However, these are
isocurvature perturbations, and so might not be related to entropy
fluctuations associated with $\Delta_i^{(s)}\ne 0$. The study
of entropy fluctuations in inflationary models is still a highly speculative
topic\cite{linde}, and its connection with initial conditions in the models
under consideration should be an exciting subject to examine in a future
paper.

\section{Saha's equation and the Jeans mass}

As mentioned earlier, it is important to compare $\tau$ and $t_{_H}$
with the timescale characteristic of the interaction rate of the photons and
electrons for Compton and Thomson scattering (the dominant radiative processes
in the temperature range of interest)

$$t_{_{\gamma}}=\frac{1}{c\sigma_Tn_e},\qquad t_{c}=\frac{m_e
c^2}{k_{_{B}}T}\,t_{_{\gamma}}\, ,
\eqno(45)$$

\noindent
where $\sigma_T \approx 6.65 \times 10^{-25}\hbox{cm}^2$ is the Thomson
scattering cross section, $m_e$ is the electron mass and $n_e$ is the number
density of free electrons, a quantity obtained from Saha's equation

$$\frac{X_e^2}{1-X_e}=\left[\frac{2\pi\, m_e\,k_{_{B}}T}{h^2}
\right]^{3/2}\frac{\hbox{exp}(B_0/k_{_{B}}T)}{n_B},
\eqno(46)$$
$$X_e\equiv \frac{n_e}{n_{_{B}}},\qquad n_B \approx n^{^{(m)}},$$

\noindent
where $X_e$ is the fractional ionization, $h$ is Planck's constant, $n_B$
is the number density of baryons and $B_0\approx 13.6$ ev is the binding
energy of the hydrogen atom. Combining (45) and (46) we obtain

$$t_\gamma
=\frac{1}{2c\sigma_{_{_B}}n^{^{(m)}}}\left[1+\left(1+\frac{4h^3\,n^{^{(m)}}
\hbox{exp}(B_0/k_{_{B}}T)}{\left( 2\pi
m_e\,k_{_{B}}T\right)^{3/2}}\right)^{1/2}
\right]. \eqno(47)
$$

\noindent
The recombination process is characterized by $X_e\approx 1/10$ in (46), so
that most free electrons have combined with protons into neutral atoms, while
the decoupling of matter and radiation strictly follows from the condition:
$t_{_{\gamma}}=t_{_{H}}$, a condition analogous to $\tau=t_{_{H}}$ and
leading to the ``decoupling temperature'' from (45) and (47) with the help
of (17b) and (22). Since $t_{_{\gamma}}$ must be smaller but qualitatively
similar to $\tau$
(increasing as
$\tau$ increases\cite{schweitzer}), a comparison between $\tau$ and
$t_{_H}$ for near decoupling temperatures should be qualitatively analogous
to that between $t_{_{\gamma}}$ and $t_{_H}$. The timescales $t_{_{\gamma}}$
and $t_c$ can be easily obtained from (45), (47) and (22) and compared with
$t_{_{H}},\,\tau$ for a set of parameters characteristic of thermodynamically
consistent models discussed in the previous section.
As shown by figure 7, displaying the ratios of $\tau,t_\gamma,t_c$ to
$t_{_{H}}$ for  $\Delta_i^{(r)}= 10^{-4}$ and $10^{-8}<\Delta_i^{(s)}< 10^{-4}$. 
For $\Delta_i^{(s)}\approx 10^{-8}$ and for near decoupling temperatures ($10^3 < T
<10^4$ K. or equivalently
$10^2<y<10^{2.6}$) we have $t_{_{\gamma}}$  smaller than $\tau$ (but of the  order
of magnitude) and overtaking
$t_{_{H}}$ along a set of values of $y$ that closely match the accepted value
$T_{_{D}}\approx 4\times 10^3\hbox{K}$. This is consistent with the 
estimation:  $\Delta_i^{(s)}\approx 10^{-8}$ obtained in the previous section (see
figure 6). On the other hand, for higher temperatures  ($T \approx 10^6$ K. or
equivalently $y\approx 1$), the relaxation parameter
$\tau$ is of the order of magnitud but greater than the Compton timescale,  the
dominant radiative process at these temperatures. It is also possible to show that
$X_e\approx 1/10$ in (46) leads to the accepted values of the recombination
temperature.

To end the discussion, we compute the Jeans mass associated to the initial
conditions of the case $\Delta_i^{(s)}\ne 0$. This mass is given
by\cite{sw},\cite{na},\cite{bor},\cite{kotu},\cite{pad}

$$M_{_J}=\frac{4\pi}{3}mn^{(m)}\left[\frac{8\pi^2
C_s^2}{\kappa(\rho+p)}\right]^{3/2}=
 \frac{4\pi}{3}\frac{c^4\chi_0\Gamma^{1/2}}{\sqrt{\rho_i^{(r)}}}
\left[\frac{\pi y^2\Psi}{3G\left(\Psi+\textstyle{{3} \over
{4}}\chi_iy\right)^2}\right]^{3/2}, \eqno(48)$$

\noindent where $\rho,\,p,\,n^{(m)}$ are given by (2), (7), (13) and (14),
$\chi_i=\rho_i^{m}/\rho_i^{r}$, $\Psi$ and $\Gamma$ follow from (19) and
$C_s$ is the speed of sound, which for the equation of state (2), has the form

$$C_s^2=\frac{c^2}{3}\left[1+\frac{3\rho^{(m)}}{4\rho^{(r)}}\right]^{-1},\qquad
\rho^{(m)}=mc^2n^{(m)}, \qquad \rho^{(r)}=3n^{(r)}k_{_B}T.$$

\noindent Evaluating (48) for $y=y_{D}\approx 10^{2.4}$, $\epsilon\approx
1/\chi_i\approx
10^3$ and $\rho_i^{(r)}\approx a_{_B}T_i^4\approx 7.5\times
10^9\,\hbox{ergs}/\hbox{cm}^3$, yields $M_{_J}\approx 10^{49}\hbox{gm}$, or
approximately
$10^{16}$ solar masses. This value coincides with the Jeans mass obtained
for baryon dominated perturbative models as decoupling is approached in the
radiative
era.

\section{Conclusions}

We have derived a new class of exact solutions of Einstein's equations
providing a  physically
plausible hydrodynamic description of a radiation-matter (photon-baryon)
interacting mixture, evolving
along adiabatic but irreversible thermodynamic processes. The conditions
for these models to
be consistent with the transport equation and entropy balance law of EIT
(when shear
viscosity is the only dissipative agent) have been provided explicitly, and
their effect on
initial conditions have been given in detail, briefly discussing the
analogy of these
conditions with gauge invariant initial perturbations in the isochronous
gauge. As far as
we are aware, and in spite of their limitations mentioned in the
introduction, we believe
these models are the first example in the literature of a self consistent
hydrodynamical
approach to matter-radiation mixtures that: (a) is based on inhomogeneous
exact solutions of
Einstein's field equations, and (b) is thermodynamically consistent. We
believe these models
can be a useful theoretical tool in the study of cosmological matter
sources, providing a
needed alternative and complement to the usual approaches based on
perturbations or on
numerical methods.

The solutions have an enormous potential as models in applications of
astrophysical and
 cosmological interest. Consider, for example, the following possibilities:
(a) {\it Structure formation in the acoustic phase}. There is a large
body of literature
on the study of acoustic perturbations in relation to the Jeans mass of
surviving
cosmological condensations. Equations of state analogous to (2) are often
suggested in this context
\cite{sw}, \cite{wei}, \cite{na}, \cite{bor}, \cite{pad}, \cite{efsta}. 
Since practically all work on this topic has been carried on with 
perturbations on a FLRW background, the exact solutions
derived and presented here may be viewed as an alternative treatment for
this problem.
 (b) {\it Comparison with Perturbation
Theory}. The models presented in this paper are based on
exact solutions of Einstein's field equations, but their initial conditions
and evolution can be adapted for a  description of ``exact
spherical perturbations'' on a FLRW background. It would be extremely
interesting, not only to compare the results of this approach with those of a
perturbative treatment, but to provide a physically plausible theoretical
framework to
examine carefully how much information is lost in the non-linear regime that
falls beyond the
scope of linear perturbations. We have studied in this paper only the case
$F=0$ in (11), thus restricting the evolution to the ``growing mode'' since all
fluid layers expand
monotonously. The study of the more general case, where  $F(r)$ in (11)
is an arbitrary
function  that could change sign, would allow a comparison with
perturbations that
include also a ``decaying mode'' related to condensation and collapse of
cosmological
inhomogeneities. (c) {\it Inhomogeneity and irreversibility in primordial
density
perturbations}. The initial conditions of the models with
$\Delta_i^{(s)}\ne 0$ are set for a hypersurface with temperature $T_i\approx
10^6$. These
initial conditions can be considered the end product of processes
characteristic of previous
cosmological history, and so the estimated value $|\Delta_i^{(s)}|\approx 10^{-8}$,
related to the
spatial variation of  photon entropy  fluctuations, can be used as a
constraint on the
effects of inhomogeneity on primordial entropy fluctuations that might be
predicted by
inflationary models at earlier cosmic time. Also, the deviation from
equilibrium in the
initial hypersurface (proportional to $[\Delta_i^{^{(r)}}]^2\approx
|\Delta_i^{(s)}|$) might be
helpful to understand the irreversibility associated with the physical
processes involved in
the generation of primordial perturbations \cite{linde}. These and other
applications are worth to be
undertaken in future research efforts.

\acknowledgements This work has been partially supported by the Spanish
Ministry of
Education and the National University of Mexico (UNAM), under grants
PB94-0718 and DGAPA-IN-122498, respectively. One of us (RAS) wishes 
to acknowledge inspiration from the feline wisdom of
Tontina, Chiquita and Pintita.

\begin{figure}
\caption{{\bf The equations $\Gamma=0$ and $\Psi=0$.}
Implicit plots of the solution of
$\Gamma=0$ (figure 1a) and $\Psi=0$ (figure 1b), in terms of
$\Delta_i^{^{(m)}}$, $\Delta_i^{^{(r)}}$ and $\log_{10}(y)$, with $\Gamma,\,\Psi$
given by (19). The grid marks the initial surface
$y=1$, and is not intersected by the surfaces $\Gamma=0,\,\Psi=0$, illustrating
that $\Gamma$ and $\Psi$ have no zeroes in the evolution range $y>1$ for density
contrasts bound by $-1\le \Delta_i^{^{(m)}}\le 1$ and $-1\le \Delta_i^{^{(r)}}\le
1$. We have considered $n_i^{(r)}/n_i^{(m)}\approx 10^9$, $m$ to be the mass of a
proton and $T_i\approx  10^6 \hbox{K}$, hence we have made the approximation (23c):
$\epsilon\approx 10^{-3}T_i\approx 10^3$. These (and the remaining) plots
were obtained with the help of the symbolic computing package Maple
V[39].}\label{f1}
\end{figure}

\begin{figure}
\caption{{\bf The ratio of relaxation vs Hubble times for the case
$\Delta_i^{(s)}=0$.}
This plot displays the ratio $\tau/t_{_H}$, as a function of $\log_{10}(y)$ and
$\log_{10}(|\Delta_i^{^{(r)}}|)$ (denoted as ``$\log(\hbox{Dr})$''), where $\tau$
is the relaxation time and
$t_{_H}=3/\Theta$ is the Hubble expansion time, given by (22) and (29)
under the condition $\Delta_i^{(s)}=0$. This ratio is less than unity for all $y$,
and $|\Delta_i^{^{(r)}}|<1$ thus violating (32b) and (32c). }\label{f2}
\end{figure}

\begin{figure}
\caption{{\bf The equations $\Phi=0$ and $\sigma=0$.}
Implicit plots of the solution of $\Phi=0$ (figure 3a) and $\sigma=0$
(figure 3b), in terms
of $\Delta_i^{^{(m)}}$, $\Delta_i^{^{(r)}}$ and $\log_{10}(y)$. The
``wall'' where the surfaces
$\sigma=0$ and $\Phi=0$ occur for values greater than $y=10^2$ corresponds
to the line $\frac{3}{4}\Delta_i^{^{(r)}}=\Delta_i^{^{(m)}}$ (or
$\Delta_i^{(s)}=0$). The plots also illustrate
that the values of $\Delta_i^{^{(r)}},\,\Delta_i^{^{(m)}}$ where a zero of
$\sigma$ occurs are
clearly distinct from those associated with a zero of $\Phi$, leading to
conditions (37). The vertical height $y=10^2$ is displayed in figures 3a and 3b,
illustrating that the values of $\Delta_i^{^{(r)}},\,\Delta_i^{^{(m)}}$ with
$10^2<y_*<10^3$ are very close to $\Delta_i^{(s)}=0$.}\label{f3}
\end{figure}

\begin{figure}
\caption{{\bf Allowed values of $\Delta_i^{^{(r)}},\, \Delta_i^{^{(r)}}$
for $\sigma=0$.} The figure displays the plane $\Delta_i^{^{(r)}},\,
\Delta_i^{^{(r)}}$ associated with figures 3. The
regions for which a zero of $\sigma$ occurs for $\Phi\ne 0$ are shown with
a gray shadow, while the blank regions are ``forbidden'' areas corresponding to
zeroes of $\Phi$. The values of $\Delta_i^{^{(r)}},\,\Delta_i^{^{(r)}}$  in the
gray areas comply with conditions (37) and so
satisfy conditions (31a), (31b) and (31c). The level curve
$\log_{10}(y)=2.4$ has been qualitatively sketched. This curve is extremely close
($\approx 10^{-8}$) to the diagonal line corresponding to $\frac{3}{4}
\Delta_i^{^{(r)}}=\Delta_i^{^{(m)}}$, marked by the letter ``{\bf A}'' (the letter
``{\bf B}'' marks the line $\Delta_i^{^{(r)}}=-\Delta_i^{^{(m)}}$. By following the
analogy with Perturbation Theory (see section 8), the line ``{\bf A}'' can be
associated with ``adiabatic'' initial conditions, while ``isocurvature'' initial
conditions would be characterized by the horizontal axis
$\Delta_i^{^{(r)}}=0$. However, the values of $y=y_{_D}$ for
the latter condition are much smaller than $y=10^{2.4}$ for all
$|\Delta_i^{^{(m)}}|\ne 0$ (see
figure 3b). Hence, initial conditions with $\Delta_i^{(s)}\ne 0$ and leading
to the right
decoupling temperature (close to the level curve $\log_{10}(y)=2.4$) cannot
be termed
``isocurvature'', but rather  ``quasi-adiabatic'' initial
conditions.}\label{f4}
\end{figure}

\begin{figure}
\caption{{\bf The decoupling temperature $T=T_{_D}=4\times 10^3$ K  for
$y=y_{_D}$.} This figure displays the implicit plot of equation (43) in terms
of $\Delta_i^{^{(m)}},\,\Delta_i^{^{(r)}}$ and $\log_{10}(y)$. The values
of $y$ for this temperature value are clearly shown to be very close to
$y=y_{_D}=10^{2.4}$.}\label{f5}
\end{figure}

\begin{figure}
\caption{{\bf The equation $\sigma=0$ for $y=y_{_D}$.}
Implicit plot of the solution to the equation $\sigma=0$ in terms of
$\log_{10}(|\Delta_i^{(s)}|)$ and $\log_{10}(|\Delta_i^{^{(r)}}|)$, denoted as
``$\log(\hbox{Ds})$'' and ``$\log(\hbox{Dr})$''. This plot illustrates the
constraints on the orders of magnitude of these quantities under the
condition that a zero
of $\sigma$ occurs for $y=y_{_D}\approx 10^{2.4}$. For values of
$|\Delta_i^{^{(r)}}|\approx 10^{-4}$, compatible with observed anisotropy of the
CBR, we have the following estimated value: $|\Delta_i^{(s)}|\approx 10^{-8}$. 
This can be a constraint on entropy fluctuations in primordial quasi adiabatic
perturbations.}\label{f6}
\end{figure}

\begin{figure}
\caption{{\bf Comparison between $\tau$, the timescales of Thomson
($t_\gamma$) and
Compton ($t_c$) scatterings and $t_{_{H}}$.} The figure depicts the ratios
$\log_{10}(\tau/t_{_{H}})$, $\log_{10}(t_c/t_{_{H}})$ and
$\log_{10}(t_\gamma/t_{_{H}}) $
along the range $0<\log_{10}(y)<3 $ for
$\Delta_i^{(m)}=(3/4)\Delta_i^{(r)}-\Delta_i^{(s)}$ with
$\Delta_i^{(r)}=10^{-4}$ and
$10^{-8}<\Delta_i^{(s)}<10^{-4}$. The curves that branch out correspond to
$\log_{10}(\tau/t_{_{H}})$ for the displayed values of
$\Delta_i^{(s)}$. The thick curves are $\log_{10}(t_c/t_{_{H}})$ (above) and
$\log_{10}(t_\gamma/t_{_{H}})$ (below) for the same values of
$\Delta_i^{(s)}$ and $\Delta_i^{(r)}$. Notice that $\tau$ is much more
sensitive to changes
in $\Delta_i^{(s)}$ than $t_\gamma$ and $t_c$. It is clear from the figure that
$\tau\approx t_\gamma$ for $\Delta_i^{(s)}\approx 10^{-8}$ for temperatures
near the
decoupling temperature, obtained from the condition $t_{_{H}}=t_\gamma$ and
closely matching
$T_{_{D}}\approx 4\times 10^3$ K. (marked by $\log_{10}(y)\approx 2.4$).
The figure
also reveals that $\tau$ and $t_c$ are of the same order of magnitude for
higher temperatures
closer to $t=t_i$. This is consistent with the fact that Compton scattering
is the dominant
radiative process in this temperature range. }\label{f6}
\end{figure}

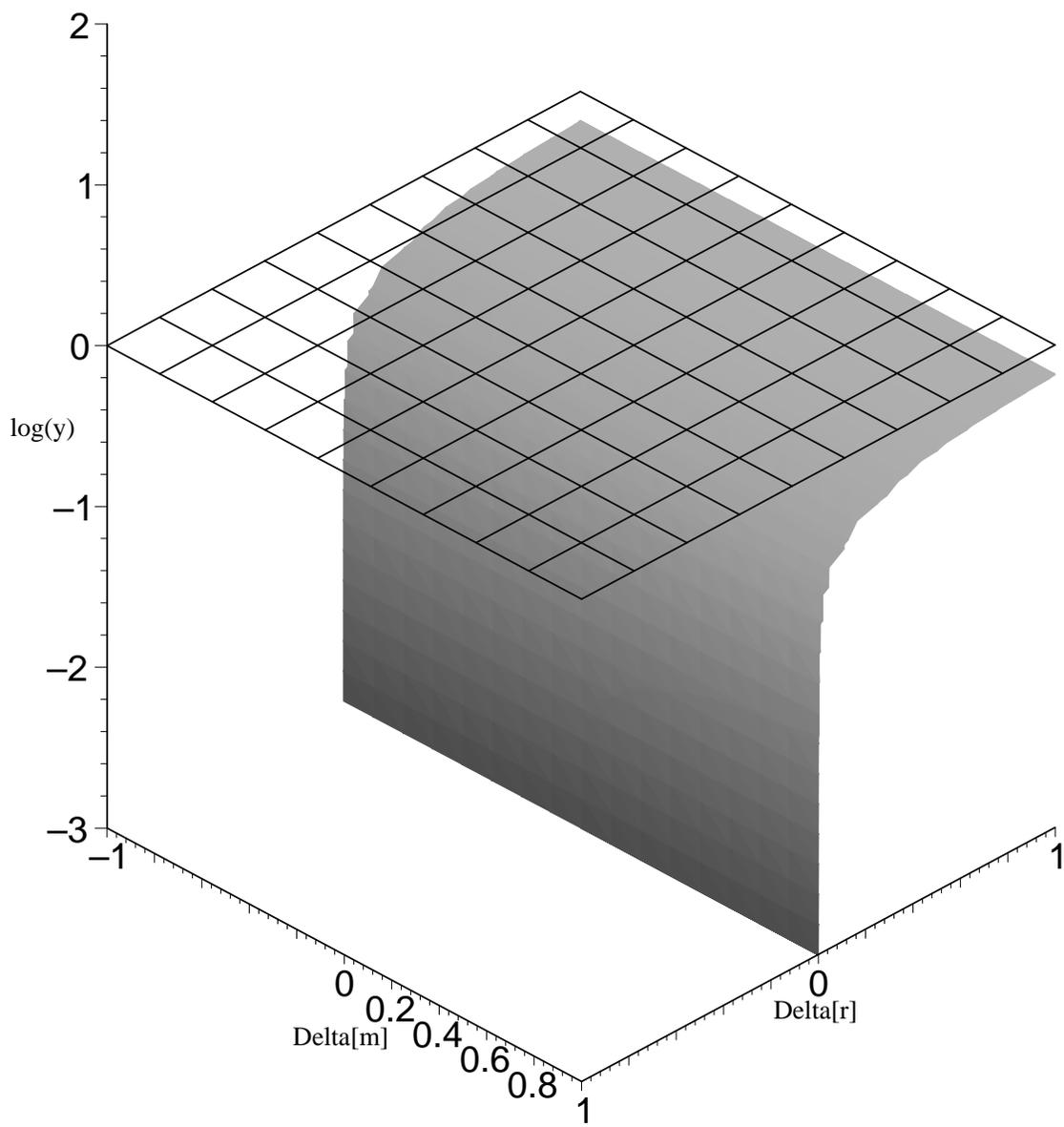
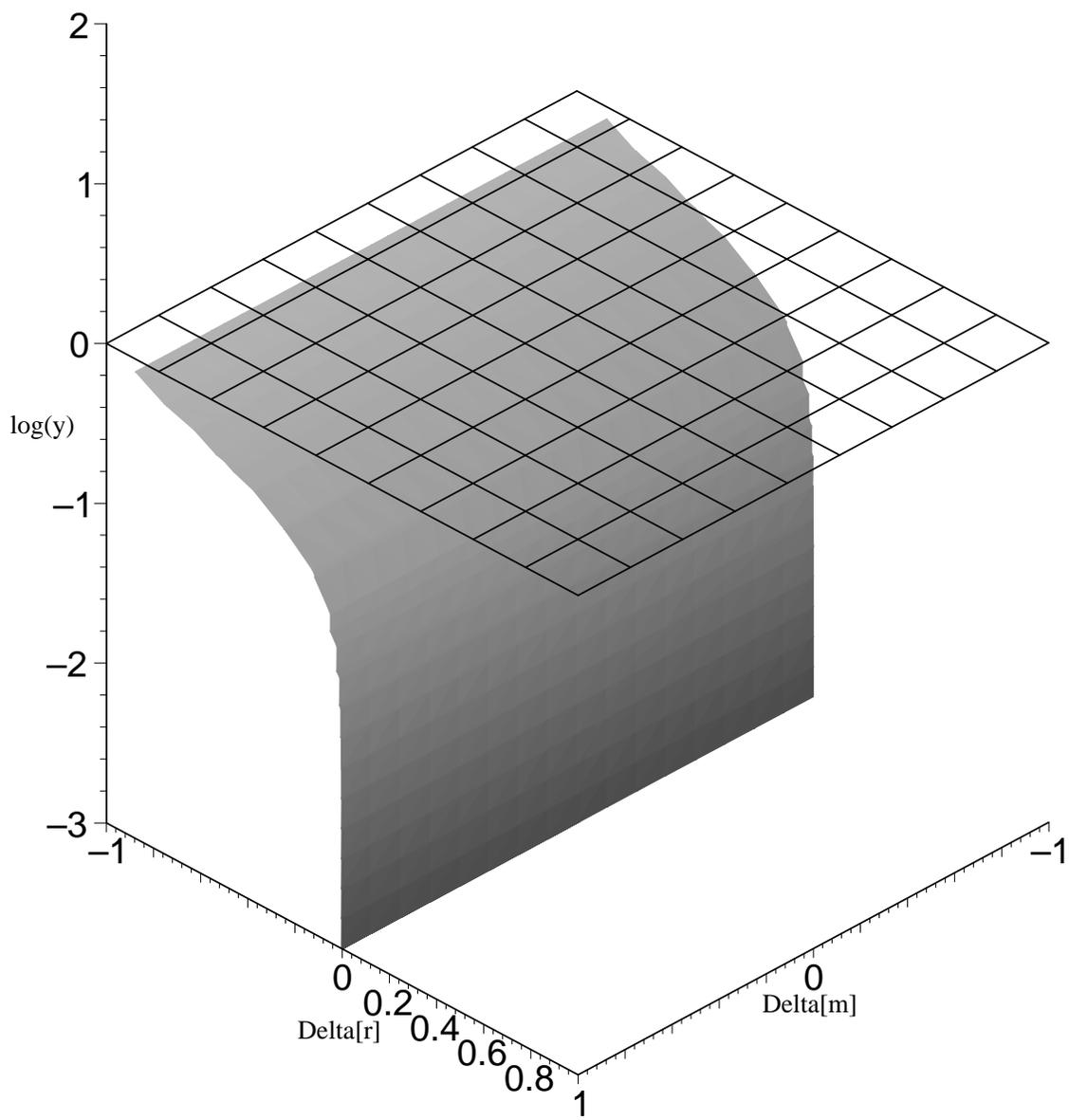
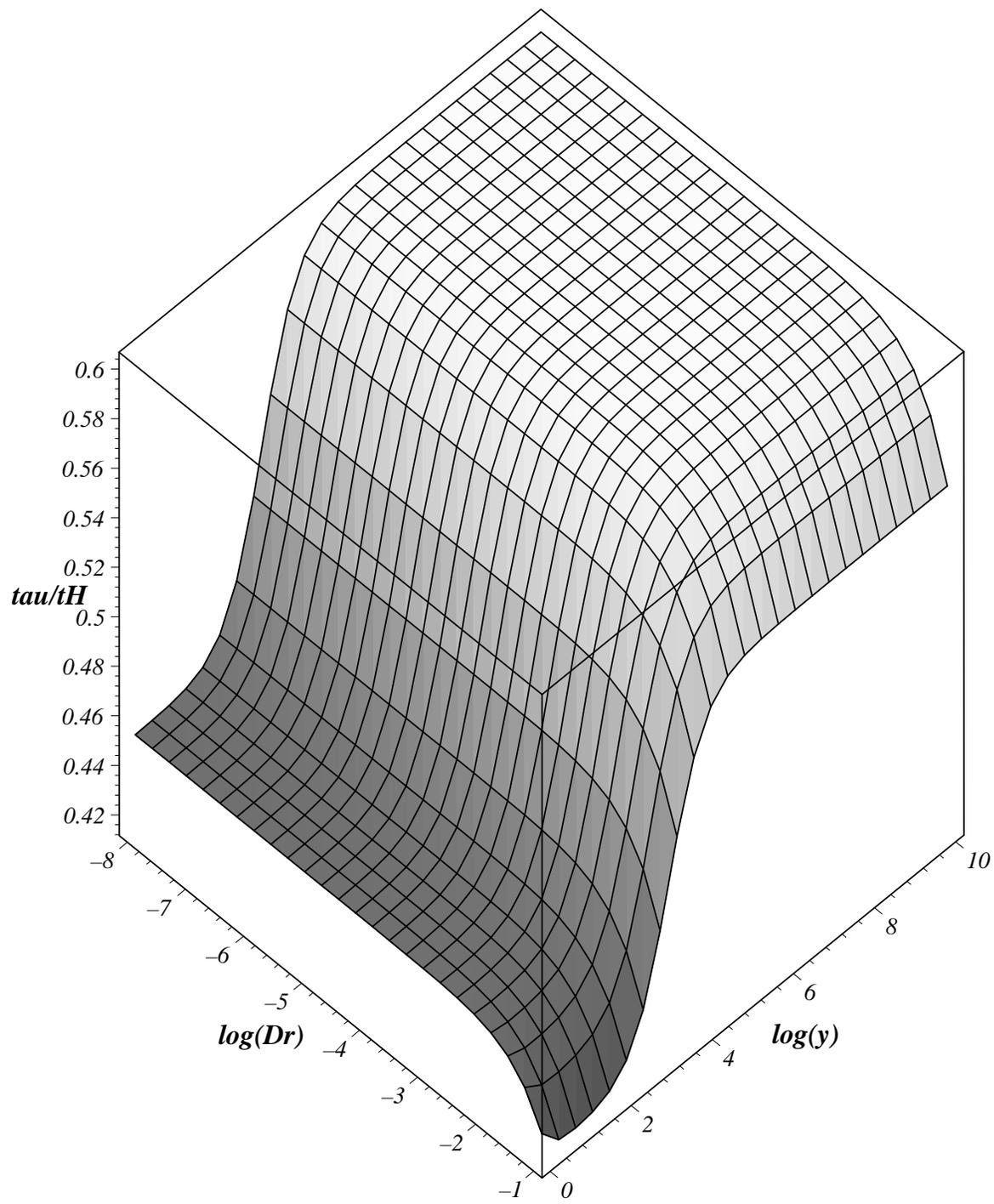
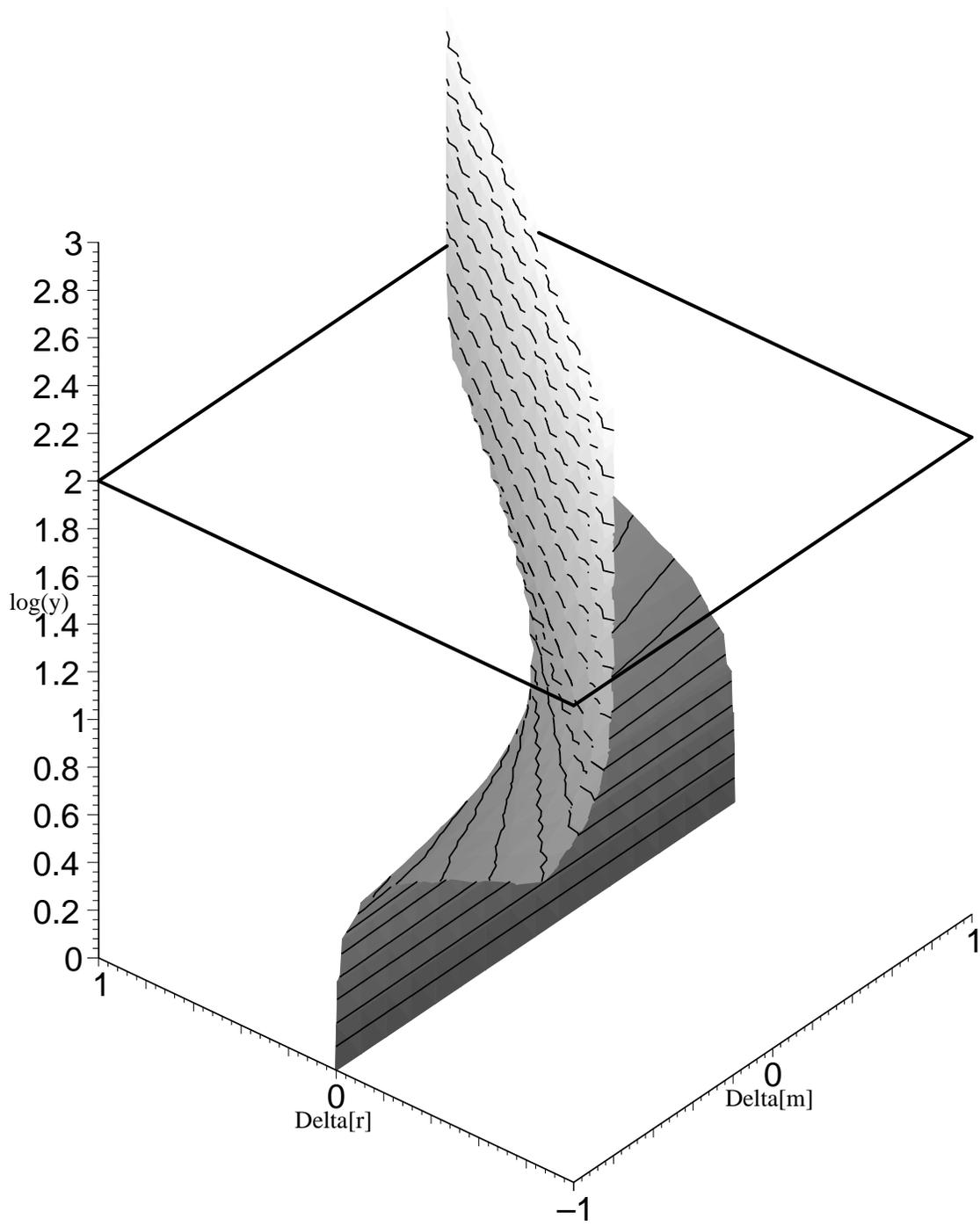
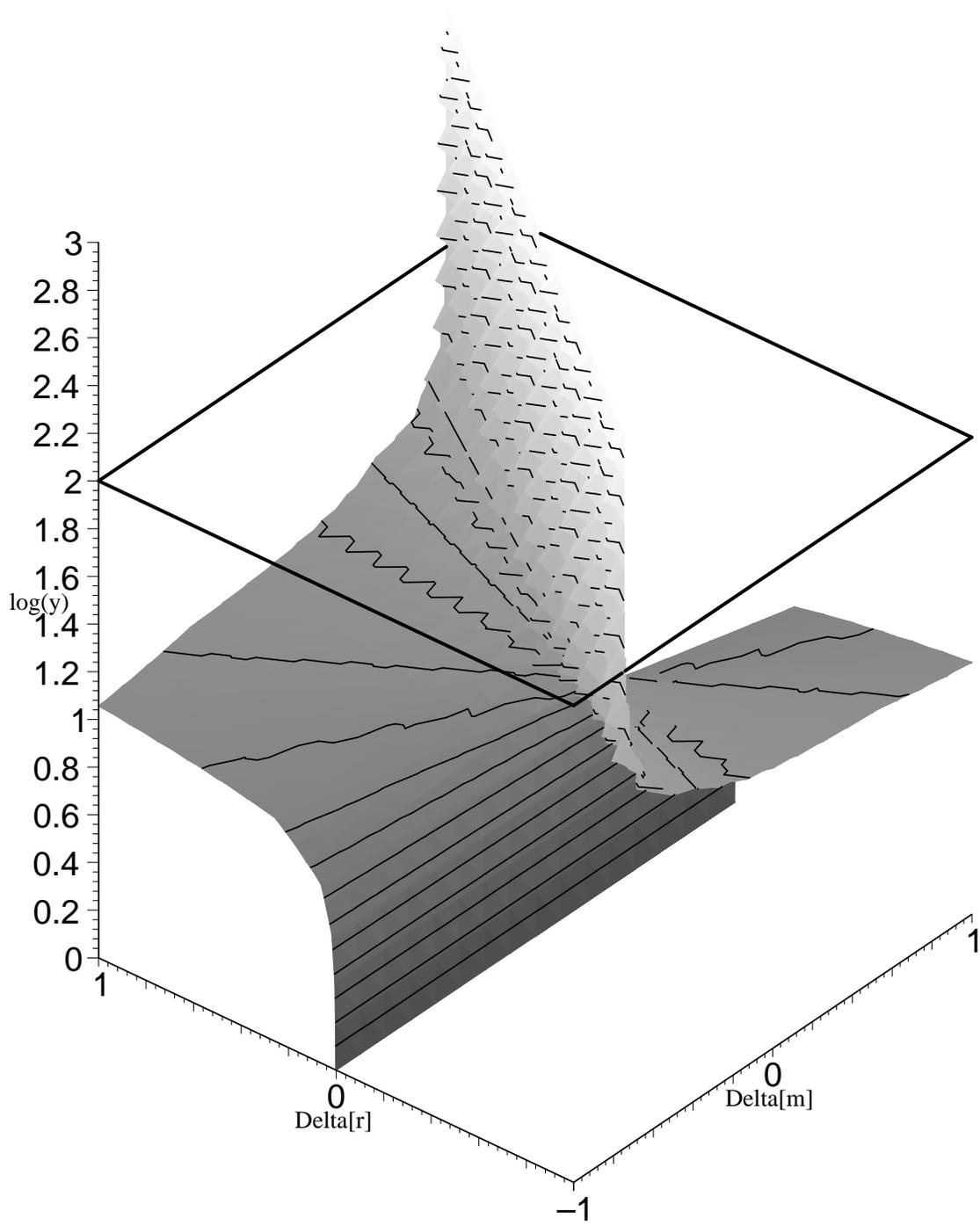
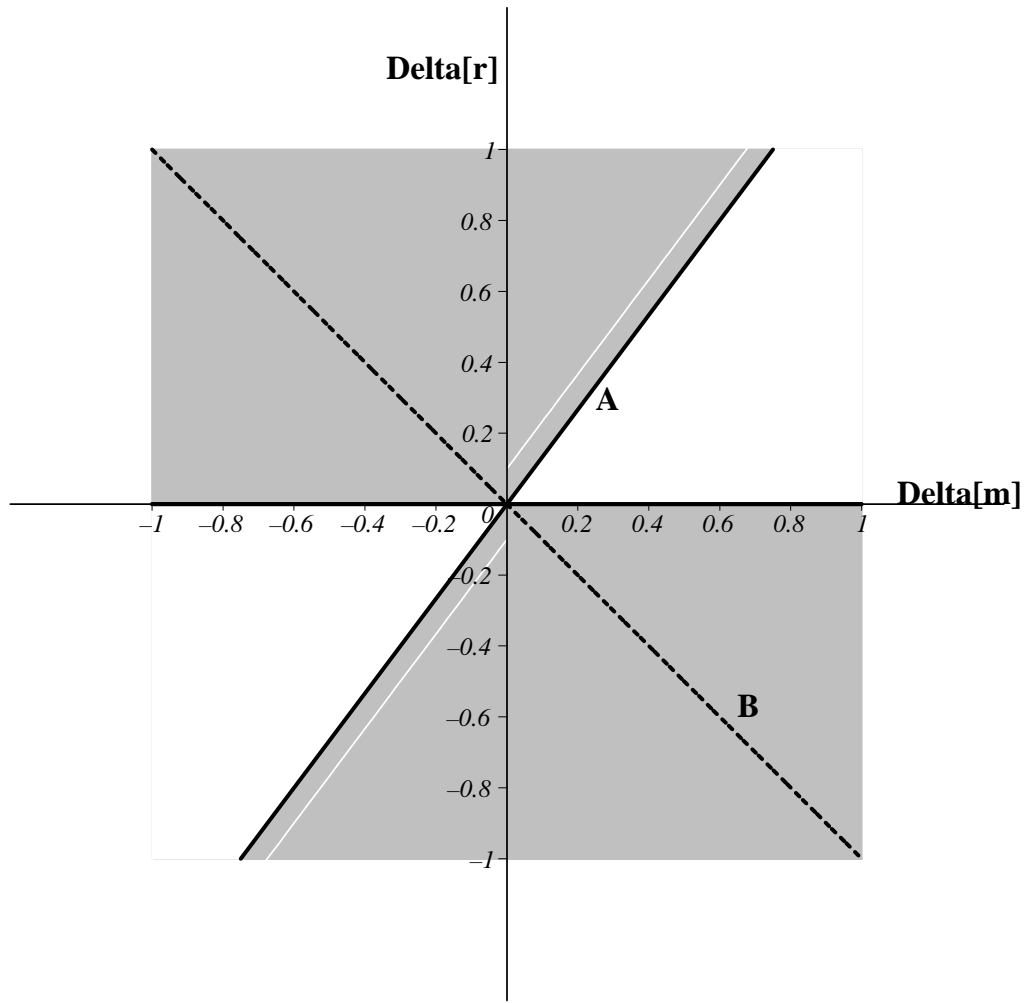
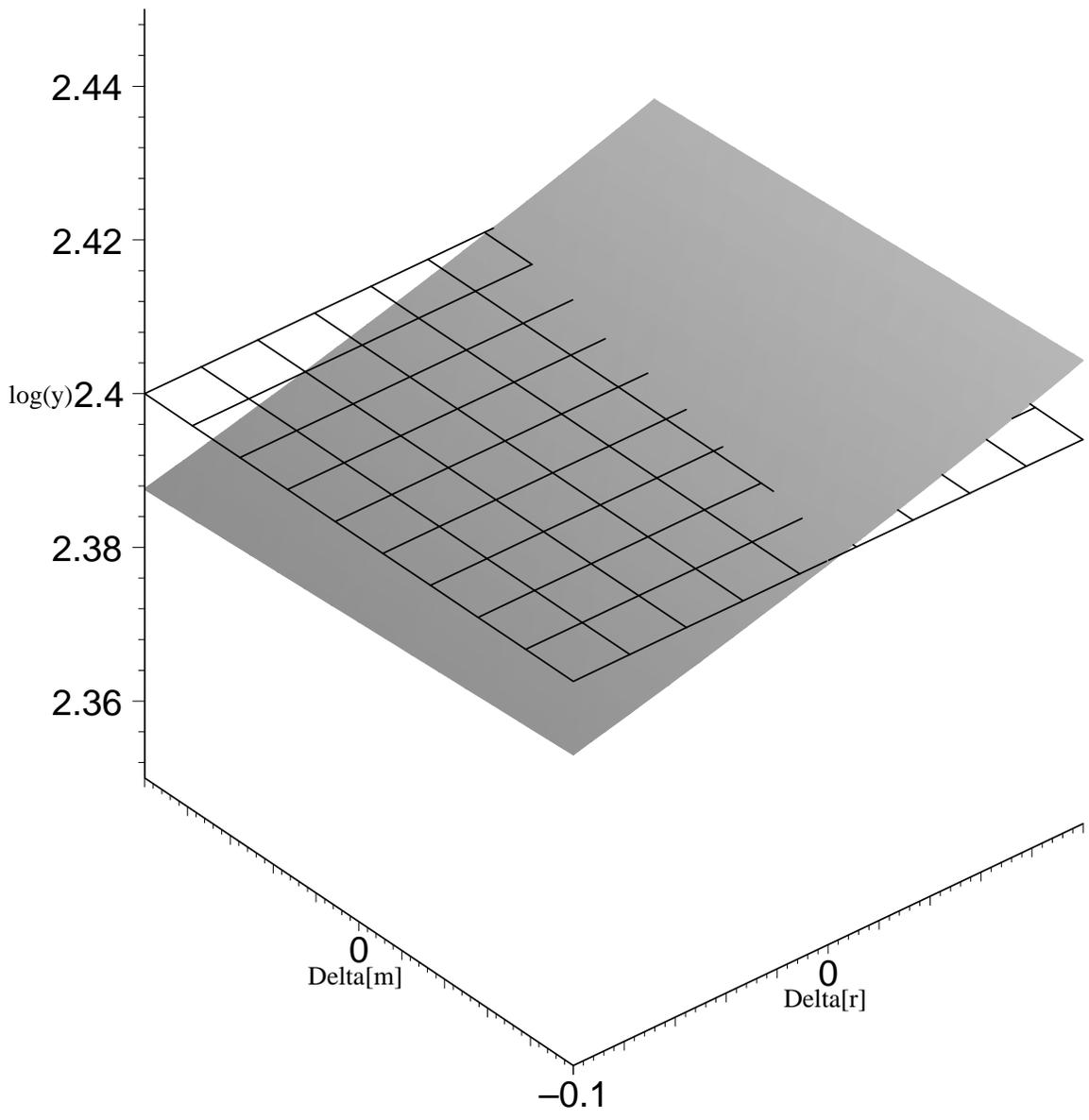
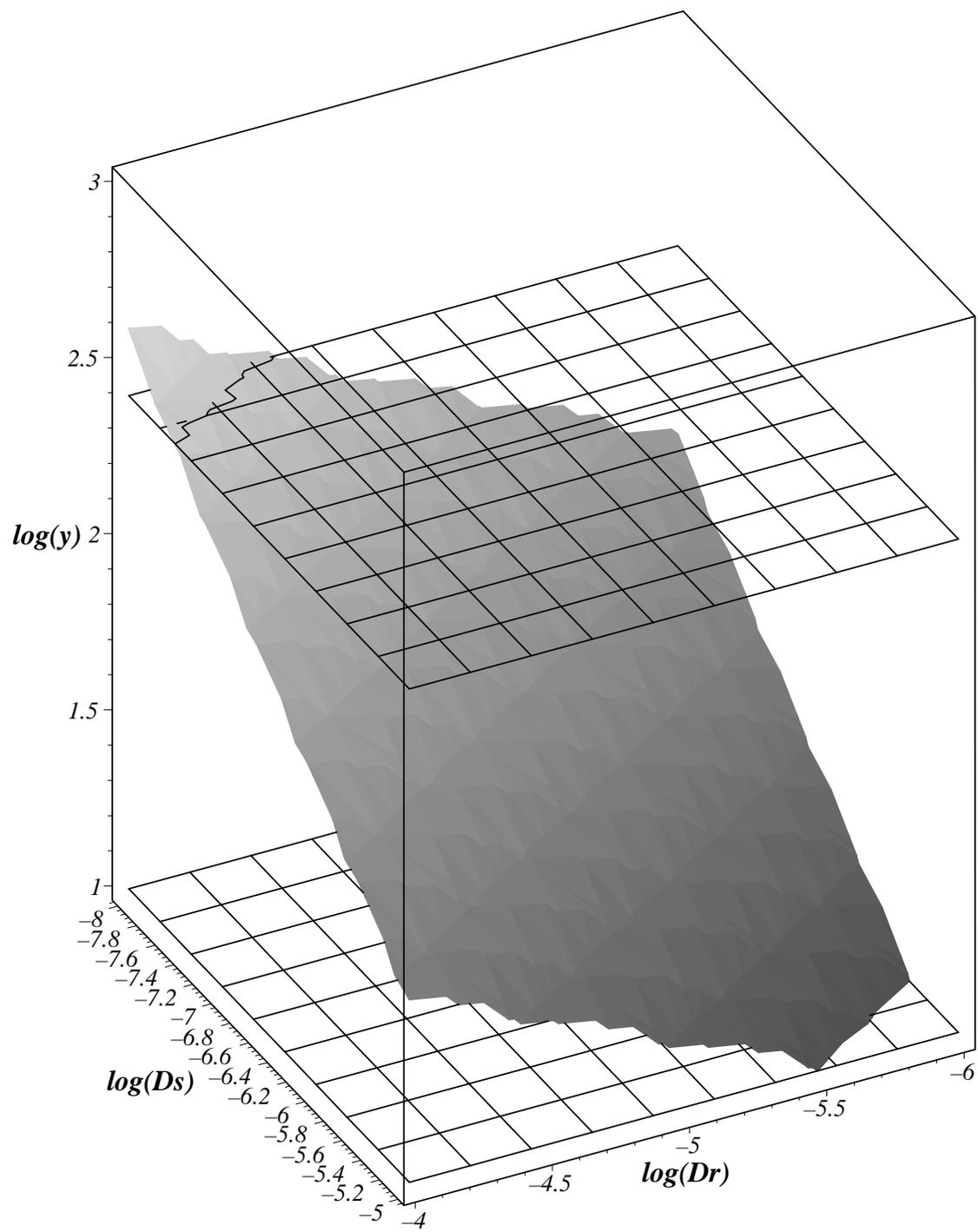
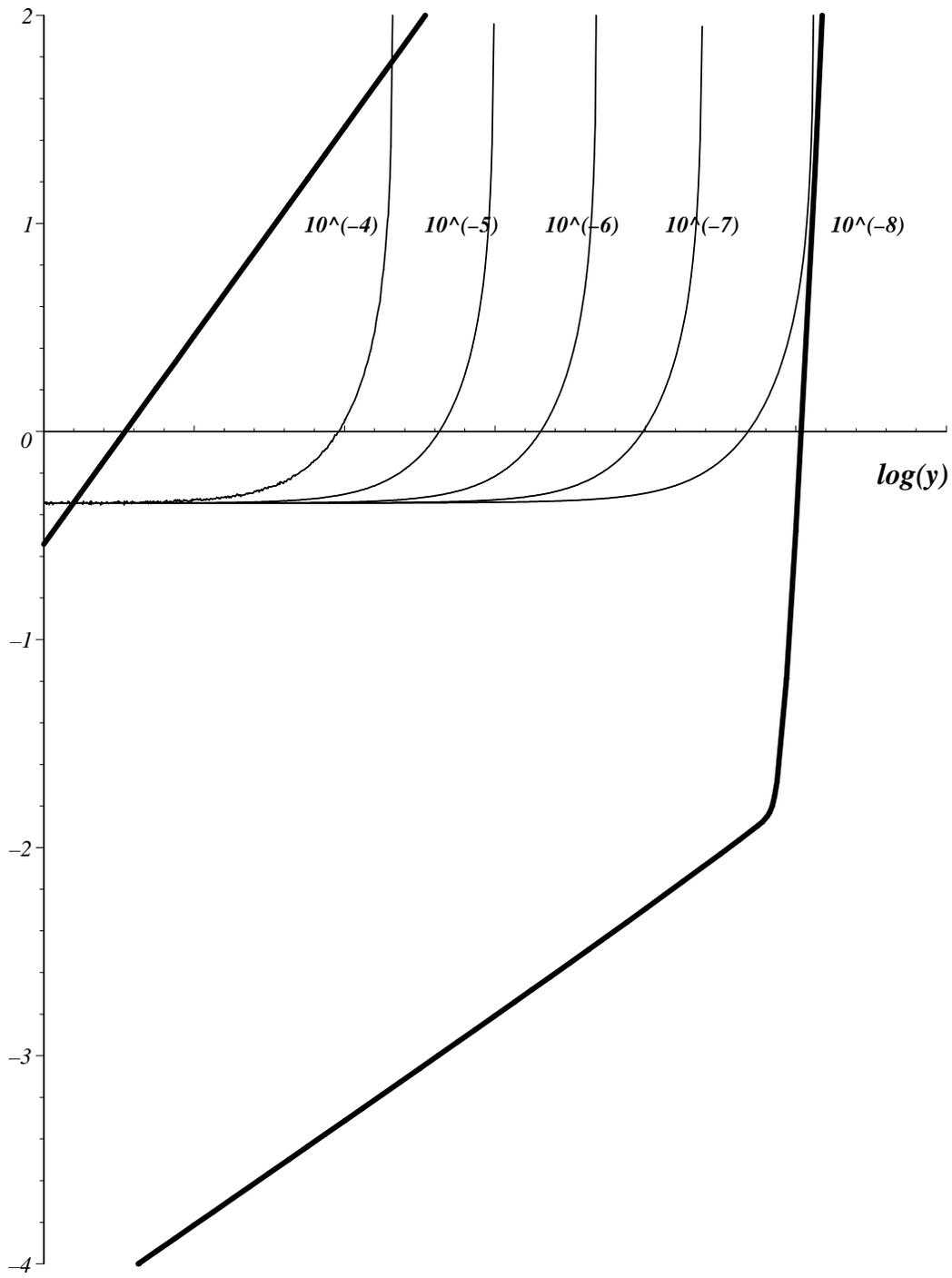
\end{document}